\documentclass[aps,prx,twocolumn,superscriptaddress,showpacs,showkeys]{revtex4-1}

\usepackage[utf8]{inputenc}
\usepackage{graphicx}
\usepackage{amsmath}
\usepackage{amssymb}
\usepackage{array}
\usepackage{multirow}
\usepackage{float}
\usepackage{color}
\usepackage{ulem}
\usepackage{caption}
\usepackage{subcaption}
\usepackage[T1]{fontenc}

\usepackage{hyperref}
\hypersetup{breaklinks=true}

\begin{document}

\title{Charge Transfer and $dd$ excitations in AgF$_{2}$}

\author{Nimrod Bachar}
\altaffiliation{Current address: Department of Physics, Ariel University, Israel }
\email{nimib@ariel.ac.il}
\affiliation{Department of Quantum Matter Physics, University of Geneva, CH-1211 Geneva 4, Switzerland}

\author{Kacper Koteras}
\affiliation{Center of New Technologies, University of Warsaw, Żwirki i Wigury 93, 02-089 Warsaw, Poland}

\author{Jakub Gawraczynski}
\affiliation{Center of New Technologies, University of Warsaw, Żwirki i Wigury 93, 02-089 Warsaw, Poland}

\author{Waldemar Trzcinski}
\affiliation{Department of New Technologies and Chemistry, Military University of Technology, gen. Sylwestra Kaliskiego 2, 00-908 Warsaw, Poland}

\author{Józef Paszula}
\affiliation{Department of New Technologies and Chemistry, Military University of Technology, gen. Sylwestra Kaliskiego 2, 00-908 Warsaw, Poland}

\author{Riccardo Piombo}
\affiliation{Dipartimento di Fisica, Università di Roma "La Sapienza", 00185 Rome, Italy}

\author{Paolo Barone}
\affiliation{Superconducting and Other Innovative Materials and Devices Institute
(SPIN),Consiglio Nazionale delle Ricerche, Area della Ricerca di Tor
Vergata, Via del Fosso del Cavaliere 100, I-00133  Rome,  Italy}

\author{Zoran Mazej}
\affiliation{Department of Inorganic Chemistry and Technology, Jožef Stefan Institute, Jamova cesta 39, 1000 Ljubljana, Slovenia}

\author{Giacomo Ghiringhelli}
\affiliation{Dipartimento di Fisica, Politecnico di Milano, piazza Leonardo da Vinci 32, 20133 Milano, Italy}
\affiliation{CNR-SPIN, Dipartimento di Fisica, Politecnico di Milano, piazza Leonardo da Vinci 32, 20133 Milano, Italy}

\author{Abhishek Nag}
\affiliation{Diamond Light Source, Harwell Campus, Didcot OX11 0DE, United Kingdom}

\author{Ke-Jin Zhou}
\affiliation{Diamond Light Source, Harwell Campus, Didcot OX11 0DE, United Kingdom}

\author{José Lorenzana}
\email{jose.lorenzana@uniroma1.it}
\affiliation{Institute for Complex Systems (ISC), Consiglio Nazionale delle Ricerche, Dipartimento di Fisica, Università di Roma "La Sapienza", 00185 Rome, Italy}

\author{Dirk van der Marel}
\affiliation{Department of Quantum Matter Physics, University of Geneva, CH-1211 Geneva 4, Switzerland}

\author{Wojciech Grochala}
\affiliation{Center of New Technologies, University of Warsaw, Żwirki i Wigury 93, 02-089 Warsaw, Poland}


\begin{abstract}
Charge transfer (CT) insulators are the parent phase of a large group of today's unconventional high-temperature superconductors. Here we study experimentally and theoretically the interband excitations of the CT insulator silver fluoride AgF$_2$, which has been proposed as an excellent analogue of oxocuprates. Optical conductivity and resonant inelastic X-ray scattering (RIXS) on AgF$_2$ polycrystalline sample show a close similarity with that measured on undoped La$_2$CuO$_4$. While the former shows a CT gap $\sim$3.4~eV, larger than in the cuprate, $dd$ excitations are nearly at the same energy in the two materials. DFT and exact diagonalization cluster computations of the multiplet spectra show that AgF$_2$ is more covalent than the cuprate, in spite of the larger fundamental gap. Furthermore, we show that AgF$_2$ is at the verge of a charge transfer instability. The overall resemblance of our data on AgF$_2$ to those published previously on La$_2$CuO$_4$ suggests that the underlying CT insulator physics is the same, while AgF$_2$ could also benefit from a proximity to a charge density wave phase as in BaBiO$_3$. Therefore, our work provides a compelling support to the future use of fluoroargentates for materials' engineering of novel high-temperature superconductors.     
\end{abstract}



\maketitle

\section{\label{sec:intro}Introduction}

Following the discovery of a high-$T_{c}$ superconductivity in the cuprate oxide (CuO) family, there has been an ongoing search for other systems in which it will be possible to replicate such novel properties. Apart from purely fundamental research into the underlying physics of the unconventional superconducting state, there has also been the goal to find $T_{c}$ at higher temperatures.

In most cases, the key players of this approach were elements originating - like copper - from the transition metal group of the periodic table, and over time several candidates came into focus such as iridates, nickelates, and vanadates. The vanadates~\cite{Cyrot1990,Viennois2010} appear to be extremely resilient to external doping~\cite{Deslandes1991} unlike the various compounds of the cuprate family. Superconductivity under doping was predicted theoretically in iridates~\cite{Wang2011}, which are isostructural to the cuprates and share several similarities with the properties of the antiferromagnetic (AFM) phase. However, there are some distinct differences compared to the cuprates. First, there is a strong competition between electronic correlations, spin-orbit coupling, and crystal field energy scales in the iridates. Second, Sr$_{2}$IrO$_{4}$ is a Mott insulator~\cite{Wang2018}, while La$_{2}$CuO$_{4}$ is a Charge Transfer (CT) insulator. Extensive experimental studies via various doping approaches did not result in any signatures for superconducting properties even upon heavy doping levels, as in the La$_{2-x}$Sr$_{x}$IrO$_{4}$ compound~\cite{Chen2015,delaTorre2015,Wang2018}. Quite interestingly, although NdNiO$_{3}$ has neither the AFM ground state nor the strong covalent bonding commonly found in cuprates, Sr doped NdNiO$_{2}$ exhibits low temperature superconductivity in an infinite layer structure as was shown recently~\cite{Li2019}.

There is, however, another approach to replace copper, and that is by staying in the same column of the periodic table of elements and by choosing silver. It was already clear from the early stages of that paradigm that silver oxide cannot become a true charge transfer insulator because of the high second ionization energy of silver and the fact that oxygen is not a sufficiently electronegative element~\cite{Tjeng1990}. As a result, AgO has a negative charge transfer energy that ends up in the Ag$^{1+}$ oxidation state without a magnetic ordering as opposed to its sibling, CuO. Fluorine is more electronegative than oxygen, therefore has deeper $2p^6$ states, which results in a positive charge transfer energy. Several fluoroargentates were found to be isoelectronic to their cuprate sibling La$_2$CuO$_4$. However, previous work showed that their magnetic ground state is not the same as, for example, in the case of Cs$_2$AgF$_4$~\cite{McLain2006} and K$_2$AgF$_4$, both being ferromagnetic~\cite{Mazej2009}, with small local structural distortions of the AgF$_6$ octahedron stabilizing an antiferro orbital ordering.~\cite{Grochala2006}

Recent calculations predict that the magnetic ground state of KAgF$_3$ and AgF$_2$ is AFM. KAgF$_3$ has an arrangement of spins within the zig–zag chain along the crystallographic c-axis direction forming an AFM with a low temperature N\'eel transition and a theoretical gap of 0.7~eV~\cite{Mazej2009,Kurzydowski2013}. However, its quasi-1D magnetic structure cannot be compared directly with the quasi-2D AFM state of the cuprates. On the other hand, AgF$_2$ has a neutral-plane stacked structure due to the inherent F$^-$ character (instead of O$^{2-}$ in CuO$_2$), although with strongly buckled planes. Therefore, AgF$_2$ is the "012" equivalent (free of charge reservoir layer) of the 214 stoichiometry in cuprates~\cite{Mazej2009,Gawraczynski2019}. The N\'{e}el temperature of AgF$_2$ ($T_N =163$~K) is half of that for La$_2$CuO$_4$ ($T_N = 325$~K). LSDA+U calculations predicted a gap of 1.5~eV to 2.5~eV in various forms of AgF$_2$~\cite{Jaron2008,Grzelak2017,Kurzydowski2017}. Hybrid DFT calculations involving orbital character confirmed the striking resemblance of the electronic structure of AgF$_2$ and its cuprate analog La$_2$CuO$_4$~\cite{Gawraczynski2019}. Furthermore it predicted also the exchange energy $J$ to be about 70~meV in AgF$_2$~\cite{Gawraczynski2019,Kurzydowski2017}, half of the $J$ value in typical cuprate compounds. A one-to-one comparison of the two-magnon excitation in Raman spectroscopy data of AgF$_2$ and EuBa$_2$Cu$_3$O$_6$ confirmed the expected $J$ of about 70~meV in AgF$_2$~\cite{Gawraczynski2019}. Although the charge transfer gap between the F $2p$ state and the Ag $4d$ state was predicted by theoretical calculations, a direct experimental verification is still lacking~\cite{Mazej2009,Gawraczynski2019}.

In this work, we study the high-energy excitations of AgF$_2$ by combining optical spectroscopy and resonant inelastic X-ray scattering (RIXS). The excitation spectrum is compared with cluster computation aided with DFT computation of parameters. We show that the optical conductivity spectrum close to the charge transfer gap resembles that of the oxocuprates. We also identify several $dd$ excitations in RIXS spectra. Our experimental observations are supported by the theoretical predictions of the electronic excitations in AgF$_2$. The striking similarity between fluoroargentate and oxocuprates suggests that the former could be a promising candidate for a future high-$T_c$ superconductor and a novel and interesting testbed for further investigations of electronic correlations in quantum materials.

\section{Methods\label{sec:methods}}

\subsection{Sample preparation}\label{sec:sample}

A 10~g batch of polycrystalline AgF$_2$ has been freshly prepared as described previously~\cite{Gawraczynski2019}. Purity of the obtained sample has been verified using powder X-ray diffraction testifying to the presence of minute amounts (ca. 1 wt \%) of diamagnetic AgF only.

To obtain a compact specimen of AgF$_2$ for optical measurements, an explosion compaction procedure was applied as described in Appendix~\ref{sec:app1}. Sample cylinder was cut into ca. 2~mm thin hard compact wafers which were fine polished inside the glovebox using diamond paper before performing any optical measurements.

Samples were loaded inside the cryostat chamber using an inert gas filled glove bag and with an active flow of gas, with a loading period of a few minutes, followed by evacuation of the sample chamber to ultra-high vacuum (UHV). No visual changes of the sample surface could be seen, testifying the successful loading without any chemical deterioration of the sample.

\subsection{\label{sec:optics}Optical spectroscopy}

The optical response of polished pressed powder AgF$_2$ samples was measured by combining infrared reflectivity and ellipsometry. The sample was installed in an UHV helium flow cryostat and data was recorded at selected temperatures from room temperature down to 8~K. The cryostat is designed to maintain a high position stability of the mounted sample during cooldown. Measurements were obtained using a near normal incident reflectivity setup in a Bruker Vertex 70V Fourier transform infrared (FTIR) spectrometer in the photon energy range of approximately 3.5~meV to 0.6~eV. Calibration spectra were obtained by measuring a gold layer deposited on top of the sample using in-situ thermal evaporation. In the photon energy range of 0.5~eV to 3.5 eV the complex dielectric function was determined using a Woollam VASE\textsuperscript{\textregistered} spectroscopic ellipsometer. The ratios of the reflection coefficients for $p$ and $s$ polarized light $\rho=r_p/r_s$, were measured at incident angles of 61$^{\circ}$ and 63$^{\circ}$ (See Appendix~\ref{sec:app2}).

The optical conductivity was calculated directly from the pseudo-dielectric function using two sets of ellipsometry data at two different angles of incidence. In addition, we have fitted the ellipsometry parameters, $\Psi_{\mathrm{ellip}}$ and $\Delta_{\mathrm{ellip}}$ to the Drude-Lorentz parametrization to obtain boundaries for our confidence in the optical conductivity spectra. We have used the far-IR reflectivity data to obtain the contribution of phonons to the spectra. The fit to the lowest energy phonon at about 30 cm$^{-1}$ along with the fit to the ensemble of all far infrared (FIR) phonons were used to extrapolate the reflectivity to zero frequency. The frequency and temperature dependence of the real part of the optical conductivity will be discussed in section~\ref{sec:results}.

\subsection{\label{sec:rixs}RIXS}

A polycrystalline sample was mounted on a copper sample holder in an inert atmosphere inside a glove box, loaded in a vacuum suitcase, and transferred to the experimental chamber maintained at $\approx5\times10^{-10}$~mbar. The F K-edge X-ray absorption (XAS) spectrum was collected in the fluorescence yield mode using $\sigma$-polarisation (normal to the scattering plane) at an angle of incidence of $75^{\circ}$. Resonant inelastic X-ray scattering (RIXS) spectra were collected at F K-edge with an energy resolution of ${\delta}E\approx0.045$~eV at a scattering angle of $150^{\circ}$ at 13 K, at I21-RIXS beam line, Diamond Light Source, United Kingdom~\cite{RIXS_Diamond}. The zero-energy transfer position and energy resolution of the RIXS spectra were determined from subsequent measurements of elastic peaks from an adjacent carbon tape. The RIXS spectrum was collected with $\pi$-polarisation (parallel to the scattering plane) at 682.2~eV at an angle of incidence of $20^{\circ}$ for 30 min. The RIXS spectrum was fitted with Gaussian lineshapes for the elastic peak and phonons, with a damped harmonic oscillator model for the bimagnon, with Gaussian lineshapes for the $dd$ excitations and charge transfer excitation, and a fluorescence model~\cite{pelliciari2016} for the emission feature peaking around 5 eV. The incident energy map was collected at an angle of incidence of $20^{\circ}$ and $\sigma$-polarisation. RIXS spectra comparison after 6 hrs of X-ray beam exposure showed only a reduction in the overall emission signal, with no noticeable difference in the intensity ratios of the inelastic features (not shown). 

\subsection{\label{sec:theory}Theory}

\begin{figure}[t]
    \includegraphics[width=0.7\linewidth]{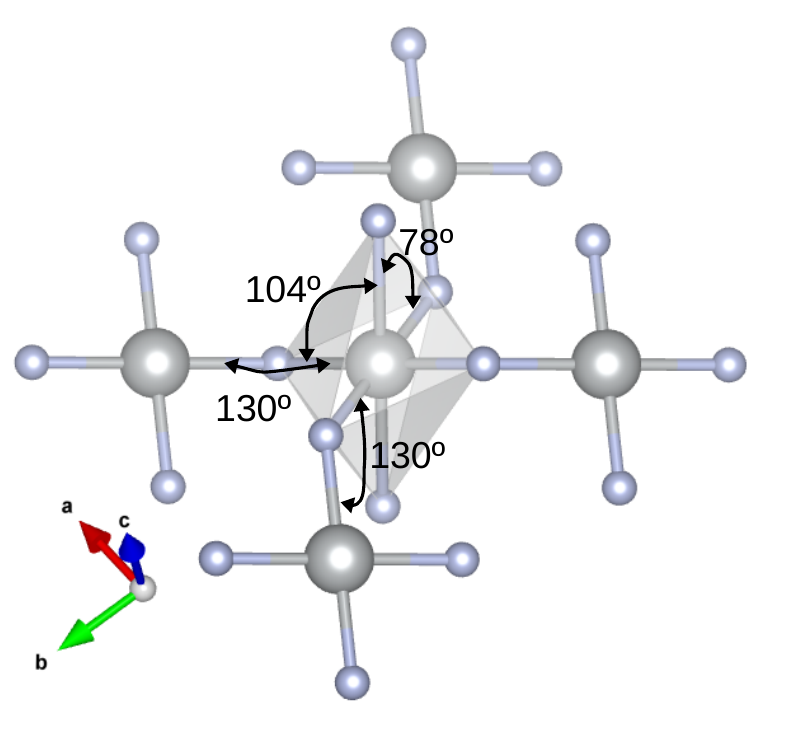}
    \caption{\label{fig:octa}The central grey octahedron shows the AgF$_6$ cluster used in the  computations. The surrounding atoms are implicitly taken into account in the definition of diagonal energies. Some key bond angles are indicated for clarity.}
\end{figure}

To model the AgF$_2$ electronic excitations with Ag in a formally $d^9$ state, we considered an (AgF$_6$)$^{4-}$ cluster reproducing the local environment of the transition metal ion as shown in Fig.~\ref{fig:octa}. The one-particle parameters were obtained from unpolarized DFT calculations of the periodic solid using the projector-augmented wave (PAW) method as implemented in VASP~\cite{Kresse1996} within the generalized gradient approximation of Perdew, Burke, and Ernzerhof~\cite{Perdew1996} (PBE) using a mesh of 8$\times$8$\times$8 $k$-points.

The band structure was obtained with the experimental lattice parameters of Ref.~\citenum{Fischer1971} and then projected onto maximally localized Wannier functions as implemented in the Wannier90 code~\cite{Marzari2012}. We used a Wannier basis with five $d$-orbitals per Ag and three $p$-orbitals per F. 
The two planar F-Ag-F bridges deviate slightly from orthogonality  ($93.3^\circ$). For the $d$-orbitals, we took the axes so that the bisectrix of the axes coincides with the bisectrix of the bridges, resulting in axes nearly oriented along AgF bonds.
For the $p$-orbitals we used the  local reference frame as in Ref.~\onlinecite{Gawraczynski2019}.

The Wannier90 one-particle Hamiltonian for the solid was truncated to the AgF$_6$ cluster yielding a Hamiltonian with 5 $d$-orbitals and $3\times6=18$ $p$-orbitals. The one-particle Hamiltonian reads 
\begin{eqnarray}
    \label{eq:htpd}
    H_{pd} &=& \sum_{\nu} \varepsilon_{d}(\nu) d^\dagger_{\nu}  d_{\nu} +\sum_{m} \varepsilon_{p}(m) p^\dagger_{m}  p_{m}  \nonumber \\
    &+&\sum_{\nu m} t_{pd}(\nu,m)  d^\dagger_{\nu} p_{m} + h.c.  \nonumber \\
    &+&\sum_{mm'} t_{pp}(m,m')  p^\dagger_{m} p_{m'} + h.c., 
\end{eqnarray}
where $\nu,m$ are spin-orbital indexes, $ d^\dagger_\nu$ creates a hole in the $d$-orbitals ($d_{3z^2-r^2}$, $d_{x^2-y^2}$, $d_{xy}$, $ d_{xz}$, $d_{yz}$) while $p^\dagger_{m}$ creates a hole in one of the 18 $p$-orbitals. We considered $pd$ hopping across the 6 F-Ag bonds [$t_{pd}(\nu,m)$] and $pp$ hopping across the 12 F-F bonds forming the edges of the octahedral cage [$t_{pp}(m,m')$]. Spin-orbit coupling was not included, so the spin is conserved.

The symmetry of the octahedra (Fig.~\ref{fig:octa}) is quite low as inversion is the only nontrivial symmetry operation allowed ($C_i$ point group). Therefore, the five $d$-levels may have different energies and hybridization matrix elements among them are allowed. Notwithstanding that, we find that with the chosen axes, off-diagonal $dd$ matrix elements on the Ag site are very small and were neglected.

Some linear combinations of $p$-orbitals are nonbonding and can be eliminated to reduce the Hilbert space. To this aim, we defined 5 symmetry adapted orbitals by the following transformation, 
\begin{equation*}
    \tilde P_\nu=\frac1{\sqrt{\tilde T_{pd}(\nu)}}\sum_m  t_{pd}(\nu,m)   p_{m}
\end{equation*}
with
\begin{equation*}
    \tilde T_{pd}(\nu)=\sqrt{\sum_m |t_{pd}(\nu,m)|^2}.
\end{equation*}
This defines a set of orbitals with maximum overlap with the $d$-orbitals. Because of the low symmetry, the resulting orbitals are not orthogonal but are easily orthogonalized, resulting in new operators $P_\nu$ and hybridization matrix elements $T_{pd}(\nu)$ which expand the same maximally hybridized subspace. The orthogonalized orbitals (hereafter $P$-orbitals) have nearly the same symmetry as the original orbitals, so they can still be labeled as $x^2-y^2$, $xy$, etc. Furthermore, they have small interorbital $PP$ and $Pd$ matrix elements between orbitals of different symmetry. We checked that keeping these matrix elements did not change the results significantly, therefore for simplicity they were also neglected.

The resulting Hamiltonian of the (AgF$_6$)$^{4-}$ cluster reads 
\begin{eqnarray}
    \label{diagonal}
    H &=& \sum_{\nu} \varepsilon_{d}(\nu)  d^\dagger_{\nu}  d_{\nu} +\sum_\nu \varepsilon_P(\nu)  P^\dagger_\nu  P_\nu \nonumber\\
    &+& \sum_{\nu} T_{pd}(\nu)\left( {d}^\dagger_{\nu}  P_{\nu} +  P^\dagger_\nu  d_\nu \right) \\
    \label{Udd}
    &+& \sum\limits_{\substack{\nu_1,\nu_2 \\ \nu_3,\nu_4}} U^{(dd)}(\nu_1,\nu_2,\nu_3,\nu_4)  {d}^\dagger_{\nu_1}  {d}^\dagger_{\nu_2} {d}_{\nu_3} {d}_{\nu_4}\nonumber. 
\end{eqnarray}

Table~\ref{DFT_par_table} shows the one-body parameters deduced from the Wannier90 computation. Setting $T_{pd}(x^2-y^2)=2.76$eV, the values of the  $T_{pd}(\nu)$ matrix elements are in very good agreement with the expressions\cite{Eskes1990} for a cluster with $D_{4h}$ symmetry using a Slater-Koster parametrization and assuming  $T_{pd}(xy)=T_{pd}(x^2-y^2)/2$ (last column of Table~\ref{DFT_par_table}). The symmetrized $P$ orbitals manifest a more evident deviation of square planar symmetry, and the $D_{4h}$ expressions for $e_P(\nu)$ are not accurate.

\begin{table}[t]
    \caption{\label{DFT_par_table} Crystal fields and hybridizations obtained from DFT and Wannier computations. All values are in eV. The diagonal energies of $P$-orbitals in Eq.~\ref{diagonal} are determined  by $\varepsilon_P(\nu)=\Delta+e_P(\nu)$. The value for $\Delta$ obtained this why is $\Delta_{DFT}=1.29$ eV but a different value can be used to study the effect of increased ionicity. The last column are the expressions for the hybridizations
    in terms of the $x^2-y^2$ matrix element and for a planar $D_{4h}$ cluster used by Eskes {\it et al.}\cite{Eskes1990}.}
    \begin{ruledtabular}
    \begin{tabular}{cccc|ccc}
                    &                   & AgF$_2$ ($C_i$)   &               &                   & La$_2$CuO$_4$     & ($D_{4h}$)                                \\  
        \hline
        $\nu$       & $\varepsilon_d(\nu)$ & $e_P(\nu)$        & $T_{pd}(\nu)$ & $\varepsilon_d(\nu)$ & $e_P(\nu)$        & $T_{pd}(\nu)$                                  \\ 
        \hline
        $z^2$       & -0.25             & 0.32              & 1.51          & 0                 & $\frac45 T_{pp}$  & $\frac{1}{\sqrt{3}}T_{pd}^\dagger$        \\
        $x^2-y^2$   & -0.28             & -0.16             & 2.76          & 0                 & $-\frac65 T_{pp}$ & ${T_{pd}^\dagger}$                        \\
        $x y$       & 0.34              & -0.05             & 1.36          & 0                 & $\frac45 T_{pp}$  & $\frac{1}{2}{T_{pd}^\dagger}$             \\
        $x z$       & 0.09              & -0.14             & 1.05          & 0                 & $-\frac15 T_{pp}$ & $\frac{1}{2 \sqrt{2}}{T_{pd}^\dagger}$    \\
        $y z$       & 0.10              & 0.04              & 1.02          & 0                 & $-\frac15 T_{pp}$ & $\frac{1}{2 \sqrt{2}}{T_{pd}^\dagger}$    \\
    \end{tabular}
    \end{ruledtabular}
    \vspace{1ex}
    {\raggedright $^\dagger$ ($x^2-y^2$) symmetry.  \par}
\end{table}

We define the CT parameter $\Delta=E(d^{10}\underline L)-E(d^9)$ where  $\underline L$ denotes a hole in the ligand and the energies are average of the indicated multiplets setting $T_{pd}(\nu)=0$. Interactions with the neighboring atoms are absorbed in the definition of $\Delta$. In the case of intracluster excitations in the insulating phase, we need to consider one hole in the cluster, making the interaction term in Eq.~(\ref{diagonal}) irrelevant. The full Hamiltonian will become relevant for intercluster excitations.

\section{\label{sec:results}Results}
\subsection{Optical Conductivity}
\begin{figure}
    \centering
    \includegraphics[width=0.8\linewidth]{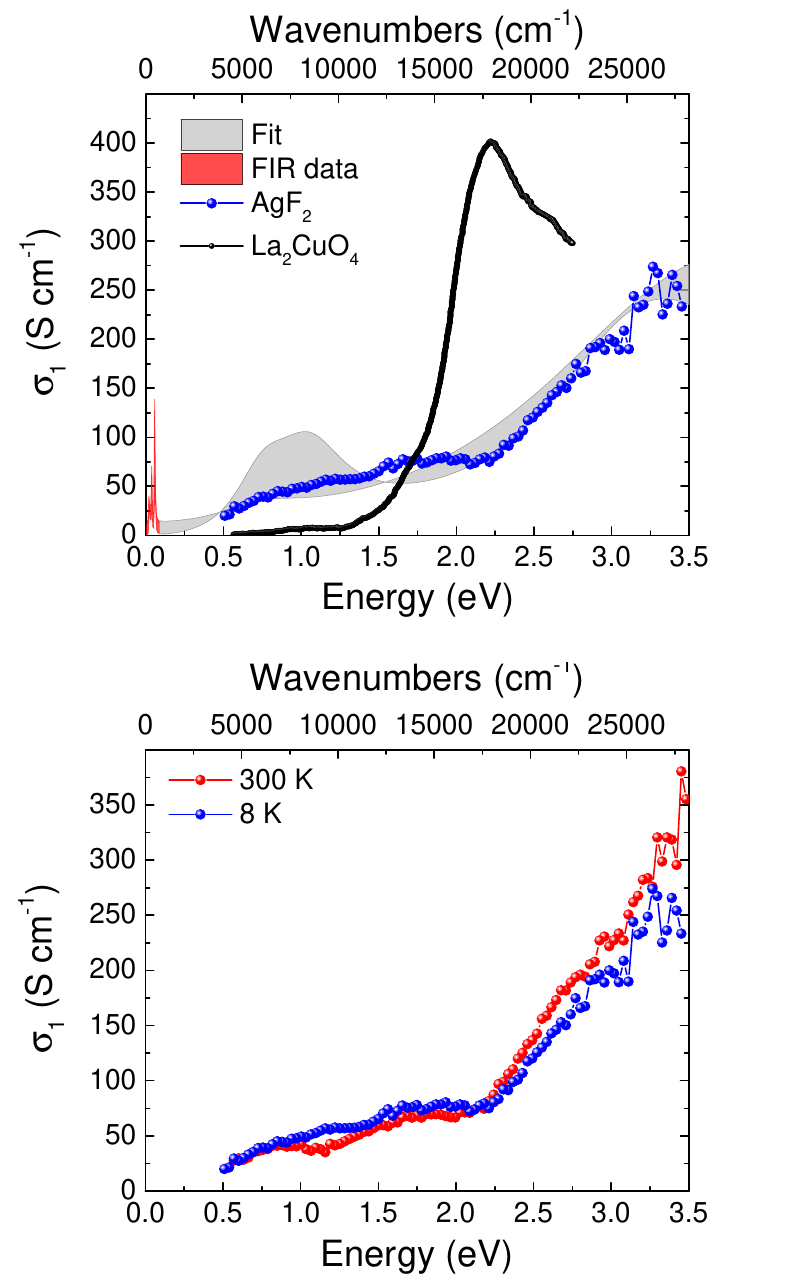}
    \caption{Real part of optical conductivity for AgF$_2$ (this work). The top panel presents $\sigma_1$ as obtained from direct calculation of the pseudo-dielectric function using ellipsometry parameters $\Psi_{\mathrm{ellip}}$ and $\Delta_{\mathrm{ellip}}$ for 2 angles of incidence as measured at a base temperature of 8~K. The shaded gray area represents the result of a fit to $\Psi_{\mathrm{ellip}}$ and $\Delta_{\mathrm{ellip}}$ data separately (see Appendix~\ref{sec:app2}) which reflects the confidence boundaries of our model with respect to the experimental data. Our data is compared with optical conductivity data adopted from Falck \textit{et al.}~\cite{Falck1992,Falck1994} on undoped La$_2$CuO$_4$. The optical conductivity in the AgF$_2$ sample shows an onset at about 1.75~eV with high-energy energy band transition which is associated with the charge transfer gap ($\Delta_{peak} \approx 3.4$~eV). The data on AgF$_2$ is compared with a charge transfer excitation peaking at $\Delta_{peak}\approx 2.2$~eV with an onset of $\approx 1.6$~eV in the equivalent oxocuprates compound La$_2$CuO$_4$. In addition, the AgF$_2$ data shows a low spectral weight sub-gap absorption suggesting
negligible doping  as intended~\cite{Falck1994}. Lower panel shows a comparison with data at 300~K showing slight modifications of the different excitations in the charge transfer sector and the sub-gap sector.}
    \label{fig:sig1}
\end{figure}

The real part of the optical conductivity, $\sigma_1$, of the AgF$_2$ sample for selected temperatures is shown in Figure~\ref{fig:sig1}. The spectra can be divided into several regimes and their equivalent excitations: far-IR phonons (seen as red sharp peaks), mid-IR absorption (above 0.1 eV), and near-IR to ultraviolet (UV) absorption. The former two low-energy excitations will be discussed in a separate publication. The high-energy spectrum (Fig.~\ref{fig:sig1}) can be decomposed into 3 major excitations. The strongest one is the high-energy absorption centered at about 3.4~eV with an onset at approximately 1.75~eV. This absorption can be associated with a strong interband transition, which sits at energies close to our experimental data range limit. 

In addition to the absorption tail of the CT excitation, we detect a broad absorption band which can be roughly decomposed into 2 modes centered at around 1~eV and 1.7~eV. As the sample is warmed up to room temperature, the 1.7~eV excitation seems to remain almost intact with respect to energy, while the 1~eV excitation seems to soften toward 0.8~eV. The two separate modes can be easily distinguished in the spectra measured at 300~K as shown in Fig.~\ref{fig:sig1}. As will be seen below, the CT excitation at 3.4~eV and the optical mode at 1.7~eV are consistent with a CT excitation and $dd$ excitations, respectively, as seen in the RIXS data. We also note a possible spectral weight transfer from the CT sector to the subgap excitations as a function of temperature and as shown in Fig.~\ref{fig:sig1}. However, the confidence boundary that we have on our optical conductivity data in this range requires a further and more precise experimental investigation of this range to understand this spectral weight interplay as a function of temperature. Further discussion of the optical conductivity spectra and comparison with that of the cuprate analog La$_2$CuO$_4$ (black line) will be given in Sec.~\ref{sec:agf2vslsco-optics}.

\begin{figure}
    \centering
    \includegraphics[width=0.85\linewidth]{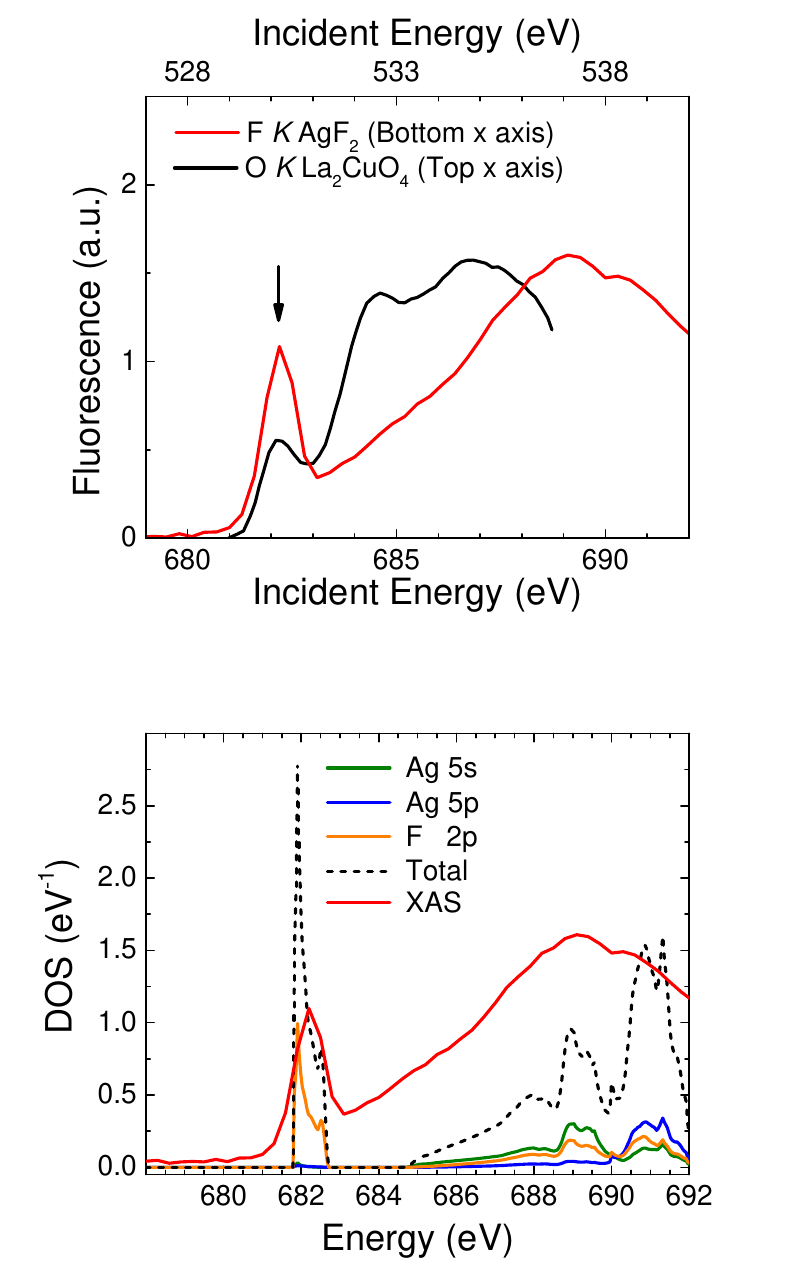}
    \caption{Top panel: Comparison of F $K$-XAS on AgF$_2$ and O $K$-XAS on La$_2$CuO$_4$~\cite{chen1994}. AgF$_2$ and La$_2$CuO$_4$ XAS energies correspond to the bottom and the top $x$-axes, respectively.
    Bottom panel: Orbital projected unoccupied DOS of AgF$_2$ from the unpolarized DFT computations. 
    The Fermi level has been shifted to 681.8~eV to facilitate comparison with F $K$-XAS (red line). The total unoccupied DOS, includes contributions from the shown symmetries above the edge as well as Ag $4d$ which will be shown later (Fig.~\ref{fig:ddos}(a)]. For the band at the Fermi level, it is nearly equal to the difference of the total and F 2p contribution.}
    \label{fig:unoccDOS}
\end{figure}

\subsection{X-ray absorption }
Figure~\ref{fig:unoccDOS} (top panel) shows the F $K$-XAS on AgF$_2$.
We attribute the edge and the sharp peak marked by the arrow (682~eV) to excitations into Ag 4$d$ orbitals via hybridization with F 2$p$ orbitals, reflecting the narrow UHB in AgF$_2$~\cite{Gawraczynski2019}. This is justified by the fact that similar $p$-$d$ hybridized peaks have been observed in F $K$-XAS of several 3$d$ metal fluorides~\cite{nakai1988,velasco2013,bondino2009}. Further support for this interpretation comes from the similarity with the O $K$-XAS on La$_2$CuO$_4$~\cite{chen1994} (black line) to be discussed in Sec.~\ref{sec:agf2_vs_lco_rixs}.

The broad absorption structure picking at 689 eV is assigned to Ag $5s$ and $5p$ states which hybridize with F-$2p$ states. These assignments are supported by the DFT computations shown in the lower panel to be discussed in Sec.~\ref{sec:dft}.

\begin{figure*}[t]
    \centering
    \includegraphics[width=0.67\linewidth]{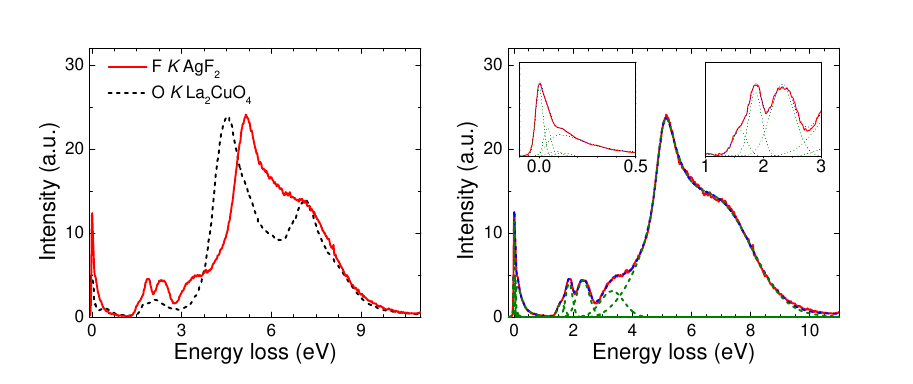}
    \includegraphics[width=0.3\linewidth]{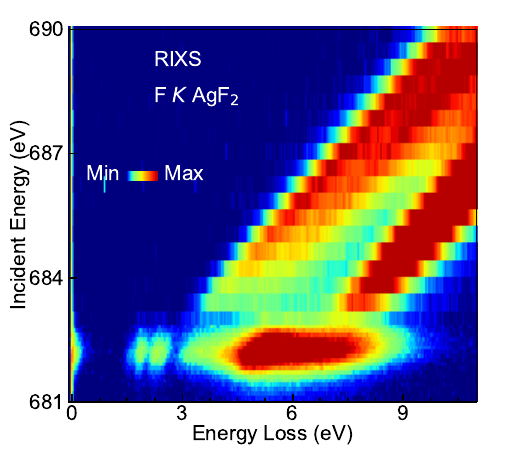}
    \caption{Left panel: Comparison of F $K$-RIXS on AgF$_2$ and O $K$-RIXS on La$_2$CuO$_4$~\cite{valentina2012}. Middle panel: F $K$-RIXS on AgF$_2$ fitted with multiple components (see text). Insets show the low-energy inelastic features and $dd$ excitations with fitted profiles. Right panel: F $K$-RIXS incident energy map on AgF$_2$ showing Raman and fluorescence features.}
    \label{fig:RIXS}
\end{figure*}

\subsection{ Resonant inelastic X-ray scattering}
Upon excitation at the F $K$ edge, a well-resolved RIXS spectrum can be observed (Fig.~\ref{fig:RIXS}). In the insets of the middle panel in Fig.~\ref{fig:RIXS}, we show that the low-energy part of the AgF$_2$ spectrum can be decomposed into an elastic peak (0~eV), phonons (0.041~eV) and their overtones and a damped bimagnon peak (centered at 0.218~eV). The energy of the assigned bimagnon peak corresponds to that observed in the Raman spectra ($\sim3J$)~\cite{Gawraczynski2019}. Bimagnons can be probed also with $K$-edge RIXS. However, to accurately determine the spectral contributions of the phonon progression and the bimagnon peak in AgF$_2$, momentum-resolved RIXS is desirable.

As shown in Fig.~\ref{fig:RIXS} (middle panel) a series of excitations at 1.59, 1.86, and 2.33~eV is observed, an assignment which will be substantiated below.  At high-energies ($>4$~eV), the RIXS spectrum is dominated by a broad feature with a strong resonance behavior as a function of the incident energy. We attribute the resonance to the excitation of CT-transitions involving an Ag site and its neighbouring F's as will be also explained below. Between the $pd$ and the $dd$ excitations, a peak is observed at 3.32 eV. The energy scale of this peak matches reasonably well with the growing optical conductivity of AgF$_2$ shown in Fig.~\ref{fig:sig1}. As such, this peak can be assigned to transitions across the CT gap from an Ag site to more distant F's and provides further confirmation of the large fundamental gap in AgF$_2$. Both the $dd$ and CT excitations are discussed in detail below while a further discussion of the comparison with the RIXS data of La$_2$CuO$_4$ will be given in Sec.~\ref{sec:agf2_vs_lco_rixs}.

The map in Fig.~\ref{fig:RIXS} (right panel) shows energy-detuned RIXS spectra collected across the F K-XAS. Energy-detuned RIXS is very useful for revealing the degree of localization of the excitations. At higher incident photon energies ($> 683$ eV), two fluorescence emission lines arise. We can view the fluorescence RIXS as an incoherent process, that is, the excitation and de-excitation are decoupled from each other. Electrons are excited from F $1s$ core level to unoccupied valence states involving Ag $5s$, Ag $5p$, and F $2p$ (c.f. Fig.~\ref{fig:unoccDOS}, bottom panel). The de-excitation is from the broad valence states to refill the $1s$ core hole resulting in a characteristic emission energy. In other words, the energy loss of fluorescence varies linearly with the incident energy. Across the resonance of the UHB peak (682.2 eV), $dd$ and CT excitations resonate at the UHB peak and show an almost fixed energy loss, i.e., Raman-like, indicating the energy loss corresponds to the energy of fixed-number-of-particle excitations (i.e. excitons, particle-hole, etc.). The localized $dd$ and CT excitations are widespread in many transition metal oxides (including oxocuprates).

\section{\label{sec:disc}Discussion}
\subsection{Optical conductivity of AgF$_2$ vs. La$_2$CuO$_4$ }
\label{sec:agf2vslsco-optics}

We now turn back to the absorption spectrum as seen in the optical conductivity data shown in Fig.~\ref{fig:sig1}. Similar behavior of a high-energy absorption was well studied, in particular in the parent compound La$_2$CuO$_4$~\cite{Uchida1991,Falck1992}. In that case, a strong peak can be seen in the real part of the optical conductivity at about 2.2~eV~\cite{Uchida1991,Thomas1992,Falck1992} as we demonstrate in Fig.~\ref{fig:sig1}. The strong peak at 2.2~eV in La$_2$CuO$_4$ is associated with the charge transfer transition between the O $p$ band and Cu $d$ band and was measured in various parent compounds of the cuprates family~\cite{Cooper1990,Thomas1992}. Taking into account the resemblance of the experimental data between AgF$_2$ and La$_2$CuO$_4$ with the supporting results of the theoretical analysis and the RIXS data, we associate the strong high-energy absorption in our data to the charge transfer excitation between the F $p$ band to the Ag $d$ band.

The broad tail down to 1.25~eV in the optical conductivity data of the AgF$_2$ sample, is similar but with a much weaker absorption to the tail in the optical conductivity data that was already reported in previous works regarding the cuprates~\cite{Thomas1992,Uchida1991,Perkins1993,Falck1994,Basov2005,Tajima2016} and as we reproduce in Fig.~\ref{fig:sig1}. Uchida \textit{et al.}~\cite{Uchida1991} demonstrated the appearance of a mid-infrared (MIR) absorption depicted as a sub-gap peak centered at about 0.5~eV in the real part of the optical conductivity of Sr doped La$_2$CuO$_4$. Falck~\textit{et al.}~\cite{Falck1994} showed that oxygen doping in La$_2$CuO$_{4+\delta}$ results in a similar MIR absorption which is dominant in sample with $\delta=0.014$ and reduced $T_N$ of $250$~K. On the other hand, in an undoped sample with $\delta=0$ and $T_N = 322~K$ the MIR absorption is negligible as can be seen in Ref.~\onlinecite{Falck1994}. 
The MIR band, which appears as a peak in the real part of the optical conductivity of doped samples, partially draws spectral weight from a higher energy range, thus diminishes the charge transfer excitation upon further doping~\cite{Uchida1991}. The intensity of the MIR band is of the order of the charge transfer excitation in nearly undoped La$_2$CuO$_4$ and even stronger for intermediate doping levels~\cite{Uchida1991,Falck1994}. Since we detect a MIR-NIR absorption band which is weak compared to that of the CT excitation, we conclude that the AgF$_2$ is practically in its undoped phase, as intended.

\subsection{AgF$_2$ vs. La$_2$CuO$_4$ from XAS and RIXS }
\label{sec:agf2_vs_lco_rixs}
In Fig.~\ref{fig:unoccDOS} (top panel) we compare the F $K$-XAS on AgF$_2$ to O $K$-XAS on La$_2$CuO$_4$~\cite{chen1994}. The peak at the absorption threshold of La$_2$CuO$_4$ reflects the O 2$p$-Cu 3$d$ hybridization and originates upon excitation to the Upper Hubbard Band (UHB) of predominantly 3$d_{x^2-y^2}$ character in this material~\cite{chen1994}. The broad structures ranging from $\sim$ 531 eV to 537 eV reflect hybridizations between unoccupied O $2p$ and La $5d$ and $4f$ DOS.
In AgF$_2$ there are no ions playing the role of La, so a different explanation for the broad absorption is needed (see below). 

As an additional comparison, O $K$-RIXS spectrum of La$_2$CuO$_4$ is also presented in Fig.~\ref{fig:RIXS} (Left panel). The low-energy excitations around 2~eV resemble the well-known $dd$ excitations of cupates~\cite{valentina2012}. The bimagnon excitation at 0.218~eV of the AgF$_2$ sample is similar to that which is observed in the O $K$-RIXS on La$_2$CuO$_4$ in the MIR region~\cite{valentina2012}. Therefore, the RIXS data of AgF$_2$ and La$_2$CuO$_4$ shows an overall remarkable similarity, which facilitates the assignment of the different features noted above and further supports the claim that silver fluorides are excellent cuprate analogues.

\begin{figure}[t]
    \centering
    \includegraphics[width=0.9\linewidth]{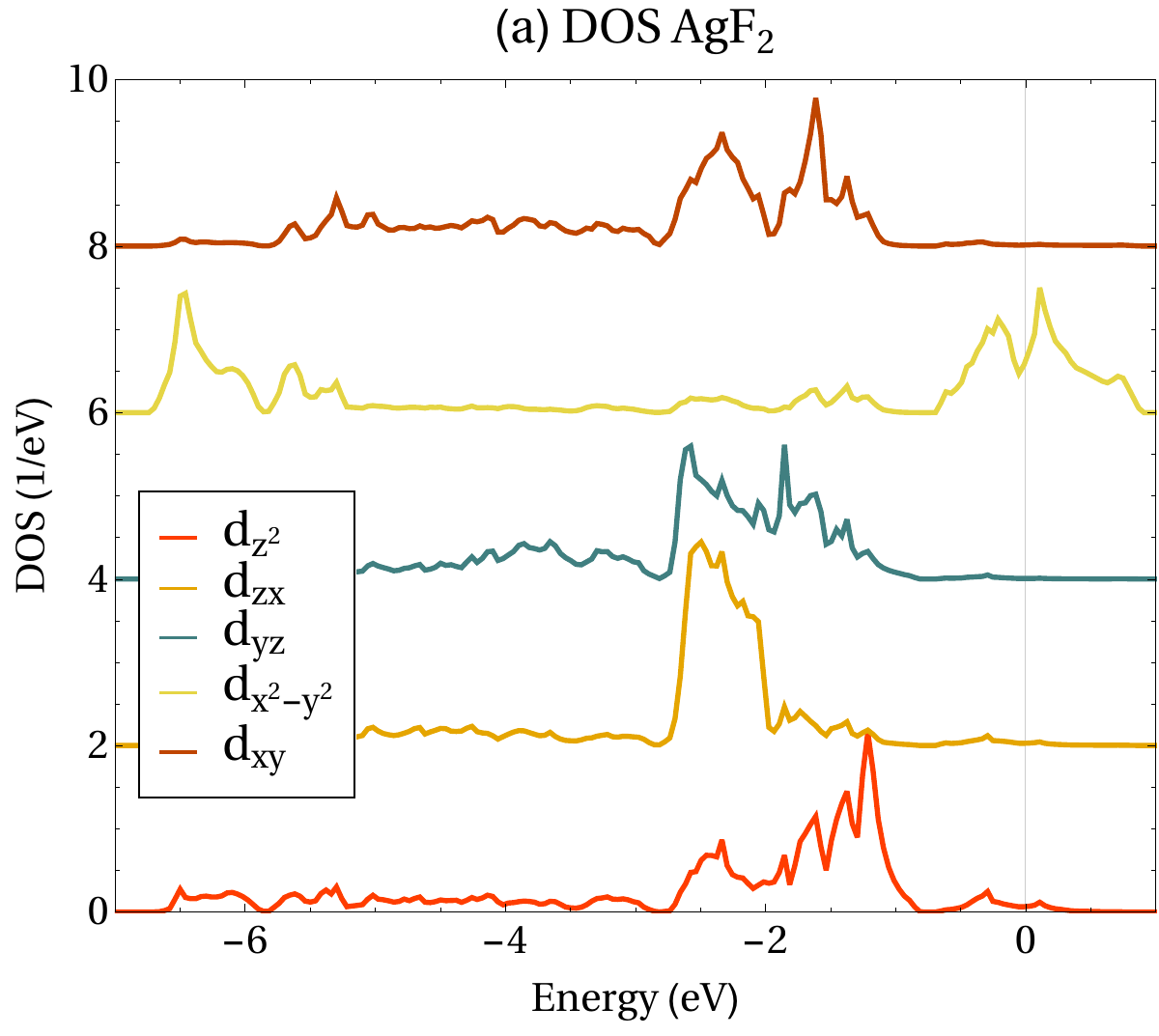}
        \includegraphics[width=0.9\linewidth]{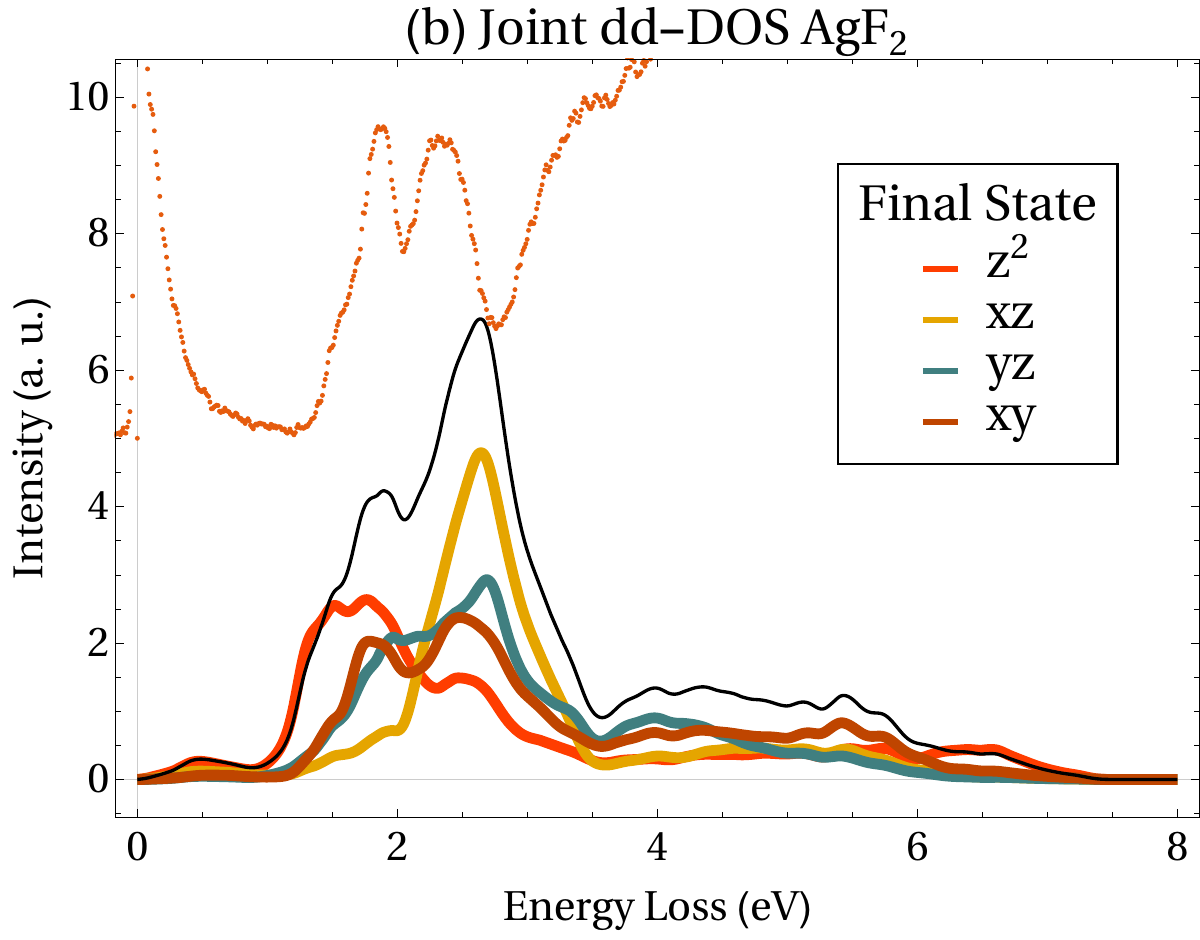}
    \caption{\label{fig:ddos} (a) DOS of AgF$_2$ projected on Ag $d$ orbitals from the unpolarized DFT computations. (curves have been shifted by 2/eV for clarity). The Fermi level is at zero energy.  
    (b) joint DOS between the unoccupied DOS of $d_{x^2-y^2}$ character and occupied DOS of different character representing $dd$ transitions of a hole from the half-filled $d_{x^2-y^2}$ to a final state (labeled by the final state) character. The thin black line is the average of the different contributions. The dotted line is the experimental RIXS data.}
\end{figure}

Additional information can be learned from the comparison of cluster calculation results between the two families which will be presented in Sec.~\ref{sec:cluster}.

\subsection{Comparison with Density Functional Theory Computations}
\label{sec:dft}
\subsubsection{X-ray absorption}
In the X-ray absorption process, a core electron is excited to the unoccupied F-states. Thus, the F-projected unoccupied DOS from DFT provides a first approximation to the spectra. Figure~\ref{fig:unoccDOS} (bottom panel) shows that there is a good match with the main structures observed. As anticipated, the DFT computations show that the broad absorption can be assigned to Ag $5s$ and $5p$ states. Notice that the F $2p$ projected DOS has large intensity at the position of both, $5s$ and $5p$ states which testify for the strong hybridization. 

The peak near 682~eV is, as explained above, attributed to the upper Hubbard band states which have mainly Ag-$4d$ character but are strongly mixed with F-$2p$ states. The 
weight of the $4d$ states in this region is approximately given by the difference between the total and the F-$2p$ DOS which is somewhat larger than the  F-$2p$ weight and again indicates strong hybridization.

\subsubsection{Assignment of $dd$ transitions}
Figure~\ref{fig:ddos} (top panel) shows the local DOS from DFT computations, projected on the different $d$-orbital symmetry in the nonmagnetic (metallic) solution with the abscissa origin set now at the Fermi energy. Due to the stronger hybridization compared to other symmetries (c.f Table~\ref{DFT_par_table}), the antibonding $d_{x^2-y^2}$ band is half-filled and is well separated from the other (filled) $d$-bands.  Panel (b) shows the joint DOS between the unoccupied $d_{x^2-y^2}$ and occupied orbitals of the other symmetries, representing $dd$ transitions with a constant matrix element. Averaging over the four possible final states (black line), one obtains a line shape surprisingly similar in position and overall shape to the $dd$ RIXS spectra, despite neglecting matrix element effects.  The theoretical line shape is broader than the experiment, which can be attributed to correlation-induced band narrowing effects absent in DFT. 
Neglecting minor differences, this analysis allows assigning the lower shoulder to transitions of the $d_{x^2-y^2}$ hole to $z^2$ orbitals, the higher peak to transitions to mainly $xz$ orbitals and the intermediate band to a mixed character. Such assignments are in general good agreement with previous works regarding fluoroargentates~\cite{Friebel1975,Monnier1991,Aramburu1992,Valiente1994,Mazej2015}. Notice, however, that in our work the peaks tend to have a mixed symmetry which can be attributed to the influence of the ligand orbitals which depart strongly from $D_{4h}$-symmetry as discussed in Sec.~\ref{sec:theory}. 

Being the compound an insulator, it may appear natural to describe it with a DFT+$U$ method\cite{Anisimov1992} and an antiferromagnetic ground state as a starting point. Instead, our computation of $dd$ transitions neglects the Hubbard-$U$. This is not important here because $dd$ excitations are charge neutral, i.e. an electron is taken from one-orbital and put in another orbital so that the Hubbard $U$ does not play an important role. The nonmagnetic computation, includes only shifts due to the ligand and crystal fields, which are the relevant ones.  Instead, a 
DFT+$U$ computation would include an additional large $U$ shift from the outset in a mean-field manner, spoiling the agreement. 


\begin{figure}[tb]
    \centering
    \includegraphics[width=0.45\linewidth]{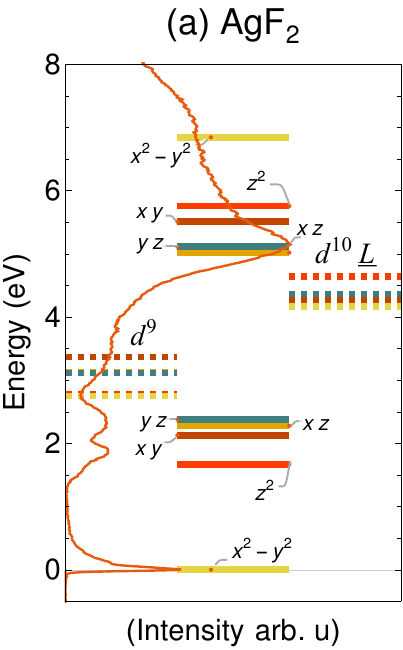}
    \includegraphics[width=0.45\linewidth]{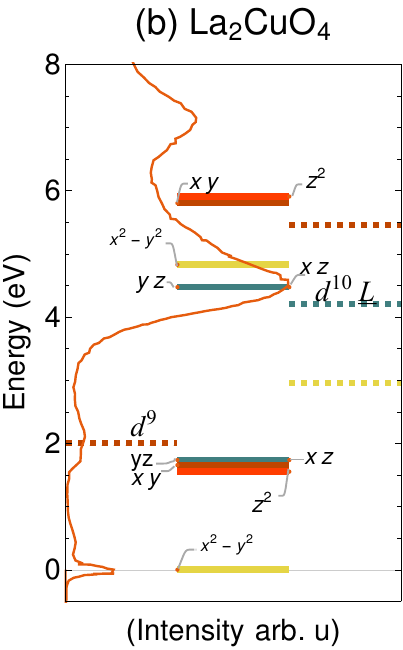}
    \includegraphics[width=0.8\linewidth]{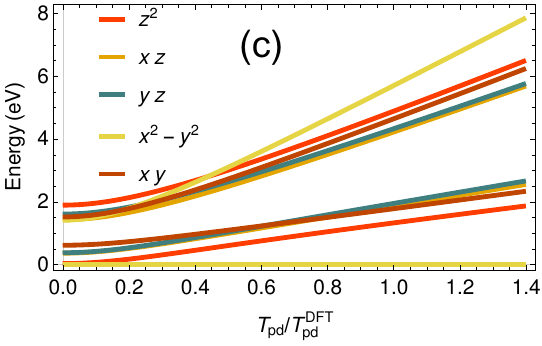}
    \caption{\label{fig:levels} Energy level diagram for one hole states in the clusters considered for (a) AgF$_2$ with $\Delta=1.29$ eV and parameters from Table~\ref{DFT_par_table} with a 20\% increase of the $T_{pd}$ matrix elements and $J=70$meV as stabilization energy.\cite{Hozoi2011} (b) La$_2$CuO$_4$ with $\Delta=2.45$ eV and $J=130$meV. 
    Dashed lines correspond to energies before hybridization in the $d^9$ configuration (left) and    $d^{10}\underline L$ (right).
    Hybridization yields bonding and  antibonding states (full lines). The ground sate corresponds to the half-filled antibonding  $d_{x^2-y^2}$ which was placed at zero energy so  the vertical scale represent energies of transitions to states of majority $d^9$ character (i.e. $dd$ transitions) around 2eV and  $d^{10}\underline L$ (i.e. charge transfer transitions) at higher energy. RIXS spectra has been superposed for comparison.
    (c) shows the Tanabe-Sugano diagram for $dd$  (lower lines set) and charge transfer transitions (higher lines set). The position of excited states is measured with respect to the $x^2-y^2$ ground state and are plotted as a function of the renormalization of the $T_{pd}$ matrix element with respect to the DFT value. The magnetic stabilization energy of the $d_{x^2-y^2}$ state mentioned in the text has been omitted.}
\end{figure}

\subsection{Comparison with cluster computations}
\label{sec:cluster}

An alternative description of the excitations, which emphasizes the local character, can be obtained with the cluster model presented in Sec.~\ref{sec:theory}. Figure~\ref{fig:levels} shows the energy levels of the one-hole configuration of the cluster for AgF$_2$ and a cuprate\cite{Eskes1990}. The dashed lines on the left (right) of each panel are the $d^9$ ($d^{10}\underline L$) configurations. Each left-right pair of a given symmetry produces a bonding and antibonding level upon hybridization shown with the full lines. Notice that because we are showing hole energies, the bonding (antibonding) states are at high (low) energies. Energies are measured with respect to the $x^2-y^2$ ground state, so the vertical scale represents the energy of transitions to the different excited states. In addition, following Ref.~\citenum{Hozoi2011} the energy of the ground state has been lowered by $J$, the magnetic stabilization energy due to the interaction with the neighbors.  We have added one time $J$ and not $2J$ since we have broken bonds with the neighbors while Ref.~\citenum{Hozoi2011} considered a cluster with 5 Cu's and excited ferromagnetic alignment of spins.

The first striking difference between cuprates and AgF$_2$ is that the $d^{10}\underline L$ configurations have much smaller splittings in the fluoride (cf. panels a and b in Fig.~\ref{fig:levels}). This is in part a consequence of smaller F-F hoppings\cite{Gawraczynski2019}. Optimized $p$ Wannier orbitals have large crystal fields splittings parametrized by $\varepsilon_p(m)$ in Eq.~\ref{eq:htpd}, but they get averaged out when projected on the $d$-symmetrized states. Notice that in the case of cuprates, $T_{pp}$ lowers the energy of the  $x^2-y^2$ ligand orbital while the hybridization with the $d_{x^2-y^2}$ rises it, resulting in a bonding $x^2-y^2$ orbital at 4.8 eV slightly above the $yz$ and $xz$ orbitals. In the case of AgF$_2$, the lowering effect of $T_{pp}$ is almost absent, so the $x^2-y^2$ bonding orbital is at much higher energies. The crossing with the other bonding levels as covalency is increased can be visualized in the Tanabe-Sugano diagram of Fig.~\ref{fig:levels}(c) (upper yellow line).

\subsubsection{$dd$ transitions.}
As a reference, we first discuss the case of cuprates for which we used one of the parameter sets considered by Eskes {\it et al.} for CuO in Ref.~\onlinecite{Eskes1990} and reproduced in  Table~\ref{par_table} (labeled as Local in the La$_2$CuO$_4$ sector). The parameter $\Delta$ corresponds to the value quoted by Eskes {\it et al.} using a slightly different definition, namely, $\Delta-T_{pp}/5=2.2$eV. From Fig.~\ref{fig:levels}(b) we see that this set of parameters gives a quite good estimate of $dd$ transition energies. As a bonus, this analysis suggests that the main structure seen in RIXS near 5 eV is a charge transfer transition from the ground state of mainly $d_{x^2-y^2}$ character to a hole in a combination of ligand orbitals with $yz$ or $xz$ symmetry. This is the lowest energy $pd$-transition within the cluster so its referred to as a "local" CT-transition.

The position and ordering of $dd$ transitions predicted in Ref.~\citenum{Eskes1990} and partially reproduced in Fig.~\ref{fig:levels}(b) was studied with the advent of RIXS two decades latter.  A detailed analysis~\cite{MorettiSala2011} of the angular dependence of RIXS matrix elements confirmed the ordering and provided a refinement of the energies. Using a more ionic parameters set with a larger fundamental gap (last column in Table~\ref{par_table}) do not produce a satisfactory agreement. We will come later to this important point.
Notice that the parameters in panel (a) have been adjusted ad hoc to fit the experiment while no such adjustment has been done for panel (b), which yields a slightly less accurate agreement for $dd$ excitations.

For AgF$_2$, we found that DFT parameters of Table~\ref{DFT_par_table} give a first ansatz for the position of RIXS structures including the DFT value for the CT parameter, $\Delta_{DFT}=1.29$ eV. However, this parameter set underestimates the energy of the $dd$ transitions which can be corrected by increasing the $T_{pd}$ matrix elements by 20\% as shown in the Tanabe-Sugano diagram of Fig.~\ref{fig:levels}(c) and panel(a). In this way, the energies of the $dd$ transitions match the experimental ones but, due to the low symmetry of the cluster, the lines should be understood as averages of the structures shown in Fig.~\ref{fig:ddos}(b) and a one-to-one correspondence of peaks and $dd$ lines is oversimplified. 
On the other hand, this analysis suggests that also for the fluoride, the main peak in RIXS near 5~eV can be assigned to the local CT transition with a final state consisting in a hole in an orbital with $xz$ or $yz$ symmetry and mainly F character [c.f. Fig.~\ref{fig:levels}(a)].

The increase of $T_{pd}$ should not be taken too seriously as it may just reflect longer range hopping process from the $d_{x^2-y^2}$ and non-magnetic stabilization terms of the ${x^2-y^2}$ ground state due to intercluster interactions beyond the magnetic correction introduced above. More importantly, increasing $\Delta$ with respect to the DFT value (i.e. increasing the ionicity) monotonously decreases the energy of $dd$ transitions (i.e., worsening the agreement). We conclude that the description of $dd$ transitions requires a quite covalent ground state in AgF$_2$. This is confirmed by the previous analysis of the joint DOS in Fig.~\ref{fig:ddos}, which is based on the same DFT computation. 

\subsubsection{Optical transitions.}

We now compare our theoretical analysis with the optical transitions. For a clean insulator, optical excitations measure the direct gap of the material. More precisely, the optical absorption may or may not show sharp transitions corresponding to particle-hole bound states (excitons), but should show an edge to a continuum of states. The threshold of the continuum corresponds to the minimum energy to separate an electron and a hole at an infinite distance and zero total momentum and defines the fundamental gap. We refer to these CT transitions as "non-local". From the optical experiments, we obtained lower and higher bounds for the fundamental gap using two different methods: i) a linear extrapolation of the edge giving a lower bound and ii) a fit with a sharp edge broadened with a Gaussian distribution giving a higher bound. We estimate $E_{gap}=1.8\pm 0.1$ eV for La$_2$CuO$_4$ and $E_{gap}=2.2\pm 0.3$ eV for AgF$_2$.

Neglecting the band formation effects, we can estimate the fundamental gap as the minimum energy to extract an electron from one cluster and add it to another cluster, i.e. 
$E_{gap}\equiv E_0(N+1)+E_0(N-1)-2E_0(N)$. This energy corresponds also to the effective $U_{\mathrm{eff}}$ in a one-band model which for cuprates provides a good description of the main charge transfer absorption band in optics\cite{Dagotto1994}. $E_0(N-1)$ corresponds to the ground state of the two-hole multiplet ($d^8+d^9\underline L$ ) 
which is the Zhang-Rice state. The $N+1$ configuration corresponds to the filled shell so there is no mutiplet but a unique state ($d^{10}$).

To compute the fundamental gap, we solved the many-body problem in the cluster using Lanczos exact diagonalization as implemented in the Quanty package\cite{Haverkort2012}. The Coulomb interaction in Eq.~(\ref{diagonal}) was parametrized in terms of Slater integrals. For Cu we used the values corresponding to the Racah parameters of Ref.~\citenum{Eskes1990}. For Ag we took the values corresponding to the Racah paramters of Ref.~\citenum{Tjeng1990} for $F^2 $ and $F^4 $ and took $F^0$ to be larger in view of the reduced screening expected in a fluoride with respect to an oxide~\cite{Gawraczynski2019}. Table~\ref{par_table} shows the resulting fundamental gap for the various parameters chosen. Both for AgF$_2$ and La$_2$CuO$_4$ the parameters that fit well the $dd$ transitions (labeled Local) correspond to $E_{gap}$ smaller than the one measured with optics. The effect is much stronger in AgF$_2$ which requires a quite small CT parameter to fit the $dd$ transitions and yet has a larger experimental $E_{\mathrm{gap}}$ than cuprates.
\begin{table}[t]
    \caption{\label{par_table} Comparison of parameter sets appropriate for local and non-local excitations and resulting fundamental gap for AgF$_2$ and La$_2$CuO$_4$. We also show the numbers of holes in the transition metal $n_d$. La$_2$CuO$_4$ parameter sets are taken from Ref.~\cite{Eskes1990}.}
    \begin{ruledtabular}
    \begin{tabular}{ccc|cc}
                        & AgF$_2$ ($C_i$)   &               & La$_2$CuO$_4$ ($D_{4h}$)  &       \\
                        &Local&Non-local & Local&Non-local \\
        \hline
        $\Delta$        & 1.29              & 2.8           &  2.45                     &  2.95 \\ 
        $T_{pd}^\dag$   & 3.31              & 2.76          &  2.3                      &  2.5  \\ 
        $T_{pp}$        & $0.11^*$          & $0.11^*$      &  1.25                     &  1.0  \\ 
        $F^0$           & 6.48              & 6.48          &  6.81                     &  7.31 \\ 
        $F^2$           & 8.19              & 8.19          & 11.41                     & 11.41 \\ 
        $F^4$           & 6.80              & 6.80          &  7.31                     &  7.31 \\ 
        $E_{gap}$       & 1.54              & 2.25          &  1.27                     &  1.8  \\ 
        $n_d$           & 0.60              & 0.73$^\ddag$  &  0.60                     &  0.66 \\ 
    \end{tabular}
    \vspace{1ex}
    {\raggedright $^\dag$ ($x^2-y^2$) symmetry.  \par
    $^*$For AgF$_2$ we used crystal field parameters from Table \ref{DFT_par_table}. For comparison we defined an effective $T_{pp}=[e_P(xz)+e_P(yz)]/2-e_P(x^2-y^2)$.\par
    $^\ddag$ Since the charge balance is determined by local transitions
    the physical value should be considered as the one computed with $\Delta$ in the Local column.\par}
    \end{ruledtabular}
\end{table}
Notice that a small $\Delta$ does not imply that the local CT transitions are at small energy. Indeed, the local $\Delta$ represents the difference in energy between the levels before hybridization (difference between left and right multiplets with dashed lines in the top panels of Fig.~\ref{fig:levels}). Instead, the local CT transition energies are set by the difference between hybridized levels (full lines in Fig.~\ref{fig:levels}). 

\subsubsection{Nearest neighbor repulsion and valence instability.}
\begin{figure}
    \centering
    \includegraphics[width=6 cm]{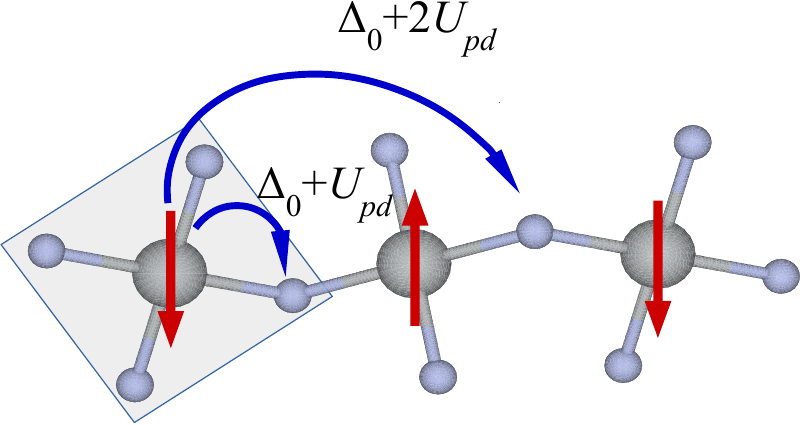}
    \caption{\label{fig:ct}Charge transfer processes within the AgF$_6$ cluster (gray region) and to a distant fluorine.}
\end{figure}

We argue that the difference in $\Delta$ needed to fit optics (labeled Non-local) and RIXS (Local) reflects interactions beyond the on-site ones considered in Eq.~(\ref{diagonal}). In particular, adding a nearest neighbor repulsion between $p$ and $d$ orbitals, $U_{pd}$, renormalizes $\Delta$ in a different way for local (intra-cluster) and nonlocal (inter-cluster) CT excitations as shown schematically in Fig.~\ref{fig:ct}. In the ionic limit, the effective $\Delta$ for local excitations ($d^9\rightarrow d^{10}\underline L$) is $\Delta^{\mathrm{loc}}=\Delta_0+U_{pd}$ while it enters as $\Delta^{\mathrm{nl}}=\Delta_0+2U_{pd}$ in the nonlocal excitations that define the fundamental gap. Here we define the difference of one-hole diagonal energies in the absence of Coulomb interactions, $\Delta_0=\bar\epsilon_p^0-\bar\epsilon_d^0$ and bar indicates average over the multiplet. 
Taking the difference of the first and second column $\Delta$ for each material, this implies a $U_{pd}\approx 0.5\pm 0.2 $ eV for cuprates (consistent with the accepted value) and $U_{pd}\approx 1.5\pm 0.5$ eV for  AgF$_2$, a value much larger than in cuprates.

In general, $U_{pd}$ matrix elements will be orbital dependent, so our estimate above should be considered as a multiplet average. Coming back to the original $p$ operators, the Hamiltonian for the nearest neighbor Coulomb repulsion in one Ag-F bond can be written as $$H_{bond}=\sum_{\nu,m}U_{pd}(\nu,m) d^\dagger_{\nu}d_{\nu} p^\dagger_{m}p_{m}.$$ We expect that also off-diagonal terms will be present, in particular 
$$H_{bond}'=\sum_{\nu\ne\mu,m}W_{pd}(\nu,\mu,m) d^\dagger_{\nu}d_{\mu} p^\dagger_{m}p_{m}+h.c.$$ in which a hole in a $p$-orbital induces a $dd$ transition. This last operator naturally explains the RIXS activity for $dd$ transitions at the ligand K-edge and can be used to evaluate the cross section. We expect that  $U_{pd}$ and $W_{pd}$ have a similar material dependence so the larger RIXS activity for $dd$ transitions in AgF$_2$ with respect to the cuprates (cf. panels (a) and (b) in Fig.~\ref{fig:levels}) is an indirect confirmation of the large intersite interactions.

So far we have used DFT to obtain spectral responses as it is customary done (Figs. \ref{fig:unoccDOS} and \ref{fig:ddos}). In general, these comparisons should be taken with a grain of salt as Kohn-Sham DFT\cite{Kohn1965} is a theory which aims to find an auxiliary non-interacting system with the same density as the interacting system, and it is not warrant that the spectral functions of the auxiliary system will match the interacting ones. In the two cases above, correlations either not play an important role or tend to cancel, which partially explain the success. 

The situation is different for the fundamental gap. The band structure of hybrid DFT computations~\cite{Gawraczynski2019} shows a smaller gap for AgF$_2$ than for La$_2$CuO$_4$ despite the fundamental gap in optics appears to have the opposite behavior. This should not be taken as a deficiency of DFT as this gap is just the gap of the auxiliary system constrained to be smaller to match the charge-distribution of the interacting system. In other words, it is a measure of the strong covalency of the interacting system, not of its gap. Indeed, model computations~\cite{Brosco2013} show that the Kohn-Sham gap in {\em exact} DFT is determined by the energy cost of neutral (i.e., local) excitations ($\Delta_0+U_{pd}$ in our case) and not the non-local ones determining the fundamental gap $\Delta_0+2U_{pd}$.
This gives further support to our finding that the DFT value for $\Delta$ has the right magnitude for describing neutral (local) transitions reconciling strong covalency~\cite{Grochala2003} in the ground state with a large fundamental gap. 

The large value of $U_{pd}\approx 1.5\pm 0.5$ eV for AgF$_2$ poses a stability problem as it implies $\Delta_0\approx -0.2 \pm 0.8$ which would make AgF$_2$ a negative charge transfer system as AgO. Taken literally and in the ionic limit, holes should populate the ligands as it occurs in formally $d^9$, silver oxide\cite{Tjeng1990}. This can be avoided if one assumes that also intra- and inter-site F-F Coulomb repulsions are present, which can stabilize the $d^9$ state. At present, the indeterminacies are too large, and it could be that $\Delta_0$ is small but positive. In any case, our results point to AgF$_2$ being close to a charge-transfer instability. In this regard, it is very suggestive that besides the magnetic brown $\alpha$-AgF$_2$ considered in this study, a metastable red-brown diamagnetic $\beta$ phase has been reported~\cite{Shen1999}, which has been interpreted in terms of a disproportionated (Charge Density Wave) ground state. Although the structure of the $\beta$ phase is not known, DFT computations~\cite{Romiszewski2007,Tokar2021} find a CDW polymorph very close in energy with respect to the usual antiferromagnetic phase, which also points to AgF$_2$ being at the verge of a charge-transfer instability.


\section{\label{sec:conc}Conclusions}

We have measured the optical conductivity and resonant inelastic X-ray scattering spectra of AgF$_2$ to study its electronic excitations. We observe a charge transfer excitation between the F $p$ bands and the Ag $d$ bands peaking at about 3.4~eV in both optical conductivity and RIXS spectra. We resolve several $dd$ excitations at 1.59, 1.86 and 2.33~eV from the RIXS spectra. We performed DFT and cluster calculations of 
the electronic structure which allowed to identify $dd$ excitations and local CT-transitions at high energy and a "non-local" CT transition determining the optical gap.

Using DFT and cluster computations, we provided estimates of the fundamental electronic parameters of this emerging quantum material, which are essential for future theoretical studies. The similarity between our data and that of the charge transfer insulator La$_2$CuO$_4$ is striking, but the subtle differences encode very interesting new physics. In particular, AgF$_2$ is predicted to be close to a charge-transfer instability due to a quite large value of $U_{pd}$. Interestingly, this parameter has been considered essential in some theories of cuprates \cite{Varma2012} so a material with an enhanced $U_{pd}$ can provide key clues to its role in determining the physics of cuprates.

Since the superconductivity in cuprates appears close to an insulating magnetic phase, while the same phenomenon in doped BaBiO$_3$ appears in close proximity of the insulating CDW phase, we expect a bright future in the search for doped\cite{Bandaru2021} and hopefully superconducting\cite{Grzelak2020} phases of AgF$_2$ which seems to combine both instabilities in the same material.  

\begin{acknowledgments}

We acknowledge J. Teyssier for his help with the ellipsometry measurements. This work was supported by the Swiss National Science Foundation through projects 200020-179157 and CRFS-2-199368. W.G. acknowledges Polish National Science Centre (Maestro grant 2017/26/A/ST5/00570). Z.M. acknowledges the financial support of the Slovenian Research Agency (research core funding No. P1–0045; Inorganic Chemistry and Technology). We acknowledge Diamond Light Source for providing the beamtime under the proposal NR24869 on the Beamline I21.
P.B., G.G. and J.L. acknowledge support from MIUR Italian Ministry for Research through PRIN Projects No. 2017Z8TS5B and No. 20207ZXT4Z. J. L. acknowledges financial support from  Regione Lazio (L. R. 13/08) under project SIMAP.

\end{acknowledgments}

\section*{Author contributions}
K.K. and J.G. contributed equally to this work. The sample was synthesized by Z.M. and shock-compressed by W.T. and J.P. N.B., K.K. and J.G. carried out the optical spectroscopy measurements. Data analysis of the optical conductivity was done by N.B. and D.v.d.M. A.N and K.Z carried out the RIXS experiment. Data analysis of the RIXS data was carried out by A.N., K.Z. and G.G. R.P. P.B. and J.L. performed DFT and cluster calculations of the electronic structure. W.G. and J.L. conceived and supervised the entire project. N.B. and J.L. wrote the manuscript with inputs and comments from all coauthors. 

\appendix

\section{\label{sec:app1}Sample preparation procedure}

\begin{figure}[t!]
    \centering
    \includegraphics[scale=.33]{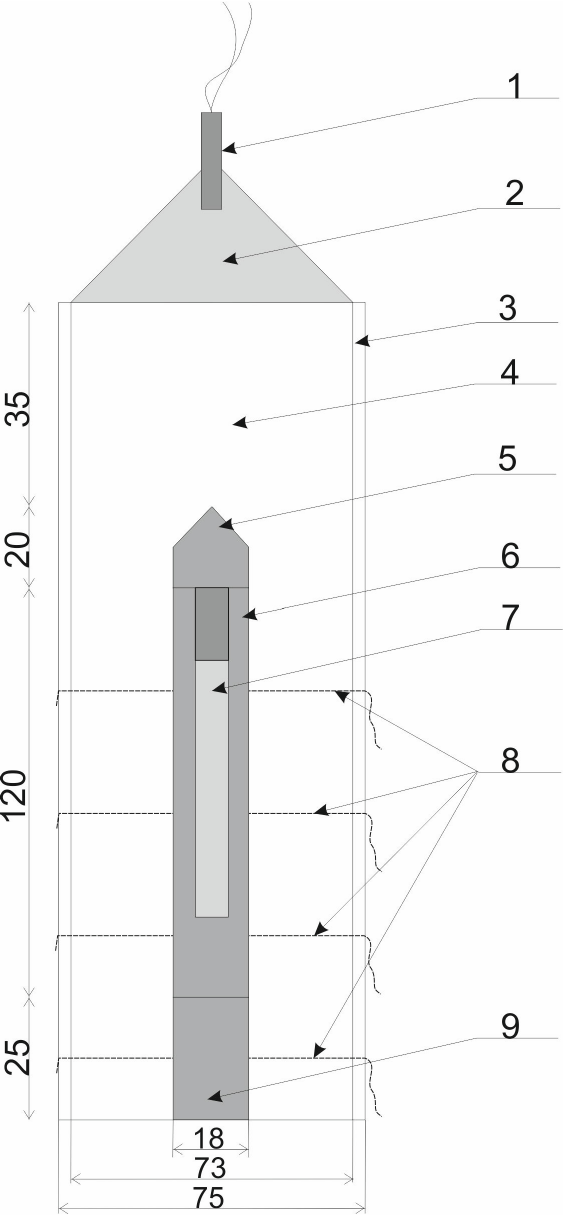}
    \caption{Experimental setup for explosive compaction. 1 – electric detonator, 2 – plane wave generator, 3 – PVC tube, 4 – explosive, 5 – upper plug together with conical hat, 6 – container with compacted sample, 7 – compacted sample, 8 – short circuit sensors for measuring the detonation velocity, 9 – momentum trap. Dimensions are given in mm on the left-hand side.}
    \label{fig:expcell}
\end{figure}

A few gram sample of AgF$_2$ has been placed inside of a 99.99\% Cu container (Figure~\ref{fig:expcell}), pressed manually using a copper cylinder, and the upper plug has been hammered into the container for even better compactness. All operations were carried out inside an argon-filled glovebox.

Explosive compaction has been performed using three distinct types of explosives; their composition and results of measurements of the detonation velocity are presented in Table~\ref{tab:expmat}. Consecutive explosions were carried out using the same container, but changing only parts 5 and 9 each time (\textit{cf.} Figure~\ref{fig:expcell}), due to their severe shock deformation.

\begin{table}[b!]
    \centering
    \begin{tabular}{| m{0.8cm} | m{4.6cm} | m{1.2cm} | m{0.8cm} |}
        \hline
        \multirow{2}{0.8cm}{Type} & \multirow{2}{4.6cm}{Explosive} & Density & $D$ \\
                &   & [g/cm$^3$] & [m/s] \\
        \hline\hline
        A   & Ammonal 2\% (98\% ammonium nitrate + 2\% aluminum powder) & 0.77 & 2600 \\
        \hline
        B   & Emulsion explosive (EmEx, mostly ammonium nitrate + fuel oil + water) & 1.17 & 5300 \\
        \hline
        C   & Emulsion explosive (EmEx) + 20\% 1,3,5-trinitro-1,3,5-triazinane (RDX) & 1.25 & 5700 \\
        \hline  
    \end{tabular}
    \caption{Explosives used for consecutive explosive compaction and their detonation velocities, $D$.}
    \label{tab:expmat}
\end{table}

Samples were undergoing explosive compaction in the following way:
\renewcommand{\labelenumi}{\Roman{enumi}}
\begin{enumerate}
    \item one sample underwent only type A explosion,
    \item one sample underwent consecutive type A and B explosions, 
    \item one sample underwent consecutive type A, B and C explosions.
\end{enumerate}

Data on high-pressure behavior of silver fluorides up to 40~GPa~\cite{Grzelak2017} were used to calculate the compression shock curve of the crystalline AgF$_2$. The equation of state of a multicomponent medium (solid-gas)~\cite{Trebinski1986} was used to describe the physical properties of the porous AgF$_2$ samples loaded by a shock wave. To estimate the temperature in the shock compressed sample, the dependence of specific heat on temperature for AgF$_2$~\cite{Gawraczynski2019} was extrapolated to high temperatures. It was assumed that the entire energy of the shock compression of a porous sample is converted into heat, which causes the sample temperature to rise. The initial density of the samples was 3.14~g/cm$^3$. It was estimated that during the type A explosion, the maximum pressure in the sample was ca. 11~GPa, the maximum temperature was ca. 1500~$\mathrm{^{\circ}C}$ and the density after loading was 5.1~g/cm$^3$. The explosion B in procedure II resulted in substantially increased maximum pressure (63~GPa), but the temperature reached only 2100~K. The density of the samples after the loading B was 5.4 g/cm$^3$. In the type C explosion (procedure III), the estimated pressure was 73~GPa and the temperature was 2000~K. Since the copper container did not explode, it was presumed that the pressure increase has substantially hindered the thermal decomposition of AgF$_2$ with the release of F$_2$ gas. Copper container was cut into pieces in an inert gas atmosphere. Indeed, visual inspection of the sample indicated that thermal decomposition and partial erosion of the container occurred only in the part of the sample adjacent to the container inner wall, while the inside of the sample cylinder was dark brown as typical for AgF$_2$. X-ray diffraction analysis revealed that the samples undergoing the procedure III were nearly pure AgF$_2$, and the apparent density of the sample was ca. 95\% of the crystallographic density; a small fraction of the sample might be amorphous.

\begin{figure}[t]
    \centering
    \includegraphics[width=0.8\linewidth]{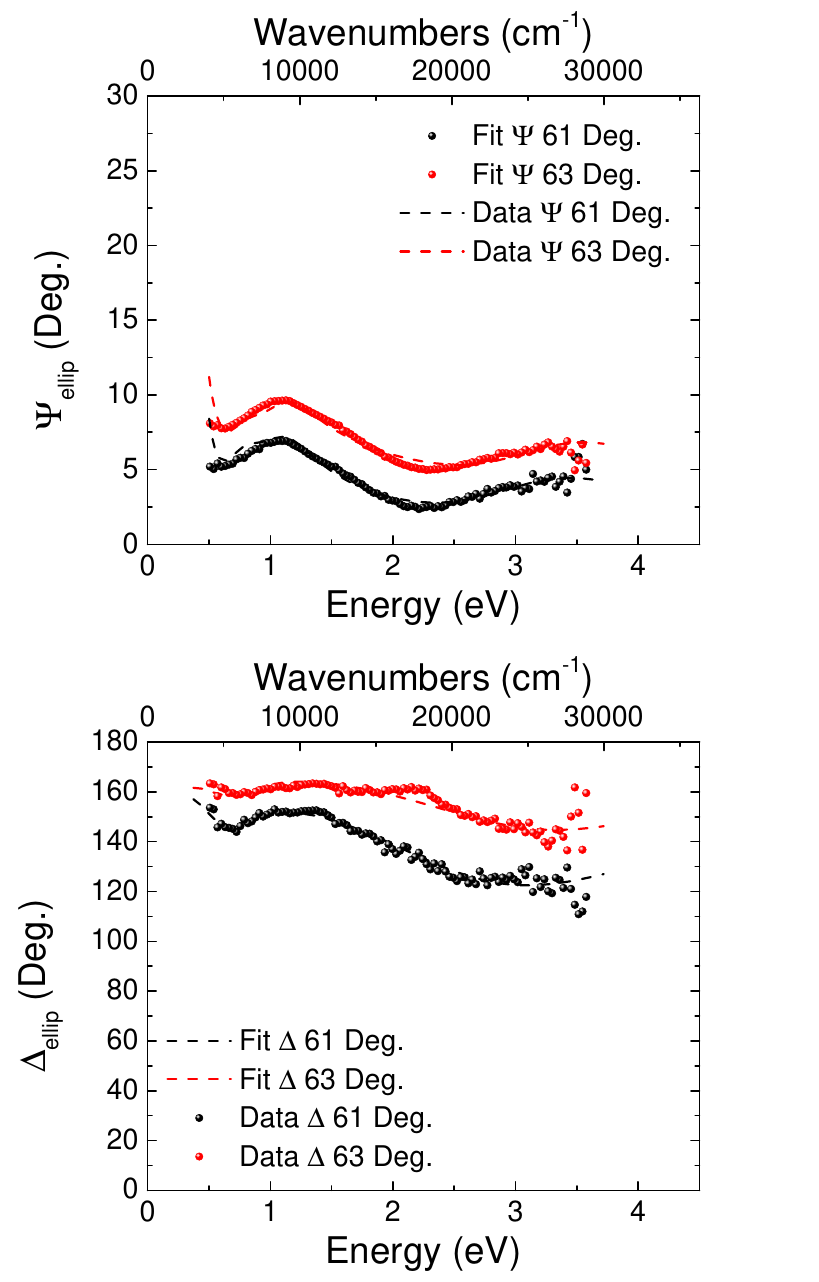}
    \caption{The ellipsometric parameters $\Psi_{\mathrm{ellip}}$ and $\Delta_{\mathrm{ellip}}$ as measured at temperature of 8~K for AOI of 61 (black spheres) and 63 (red spheres) degrees. The corresponding model fit as explained in the appendix is shown in dashed  lines.}
    \label{fig:ellipso}
\end{figure}

\section{\label{sec:app2}Ellipsometry data}

We measured the complex dielectric function using a Woollam VASE\textsuperscript{\textregistered} spectroscopic ellipsometer in the energy range from 0.55~eV to 3.5~eV. The reflectivity ratio for $p$ and $s$ polarization is defined as
\begin{equation}
	r_{p}/r_{s} = \rho = \tan(\Psi_{\mathrm{ellip}})e^{i\Delta_{\mathrm{ellip}}}
	\nonumber
\end{equation}
where $\Psi_{\mathrm{ellip}}$ and $\Delta_{\mathrm{ellip}}$ are the parametric amplitude and phase difference components of the complex reflectivity $\rho$. We measured the $r_{p}/r_{s}$ ratios at incident angles $\theta$ of 61$^{\circ}$ and 63$^{\circ}$ at selected temperatures of 300~K, 250~K, 226~K, 200~K, 176~K, 150~K, 100~K, 50~K and 8~K. The corresponding $\Psi(\omega)$ and $\Delta(\omega)$ spectra for 2 angles of incidence measured at the lowest temperature are displayed in Figure~\ref{fig:ellipso}. 

\par 

We fitted $\Psi(\omega)$ at two different angles of incidence (AOI) simultaneously following a similar but a separate fit to the $\Delta(\omega)$ spectra at two different angles of incidence using a Drude-Lorentz parametrization of $\epsilon(\omega)$. For the powder pressed sample of AgF$_2$, we have used an isotropic model for $\epsilon(\omega)$. The confidence boundaries presented in Fig.~\ref{fig:sig1} of the main text represent the possible range of $\sigma_1$ values between the two limits, i.e. a fit to $\Psi_{\mathrm{ellip}}$ data (2 AOI) and a fit to $\Delta_{\mathrm{ellip}}$ data (2 AOI), thus reflecting our error bar for the reported results. A simultaneous fit to both $\Psi_{\mathrm{ellip}}$ and $\Delta_{\mathrm{ellip}}$ was not possible to obtain under the use of a simple one layer model of $\epsilon(\omega)$ for the AgF$_2$ sample. We assume that the discrepancy is due to the polycrystalline nature of the sample. Nevertheless, the impact of this discrepancy is mostly shown around the 1~eV absorption while the CT excitation is robust for both analysis procedures. 

\par 

To extract the optical conductivity directly from the $r_{p}/r_{s}$ ratio, we have used the following pseudo-dielectric function:
\begin{equation}
    \epsilon_{ps} = \sin^2(\theta)\left[1 + \tan^2(\theta)\left( \frac{1-\rho}{1+\rho}  \right)^2 \right]
    \nonumber
\end{equation}
where $\sigma=-i \nu 2 \pi c \epsilon_0 (\epsilon_{ps}-1)$ is the optical conductivity in units of S/cm as shown in Fig.~\ref{fig:sig1} for the real part. Here $\nu$ are wavenumbers in cm$^{-1}$, $c$ is the speed of light in vacuum in m/s and $\epsilon_0$ is the vacuum permittivity in F/m.

\bibliographystyle{apsrev4-1}
\bibliography{AgF2_bib}

\begin{thebibliography}{57}%
\makeatletter
\providecommand \@ifxundefined [1]{%
 \@ifx{#1\undefined}
}%
\providecommand \@ifnum [1]{%
 \ifnum #1\expandafter \@firstoftwo
 \else \expandafter \@secondoftwo
 \fi
}%
\providecommand \@ifx [1]{%
 \ifx #1\expandafter \@firstoftwo
 \else \expandafter \@secondoftwo
 \fi
}%
\providecommand \natexlab [1]{#1}%
\providecommand \enquote  [1]{``#1''}%
\providecommand \bibnamefont  [1]{#1}%
\providecommand \bibfnamefont [1]{#1}%
\providecommand \citenamefont [1]{#1}%
\providecommand \href@noop [0]{\@secondoftwo}%
\providecommand \href [0]{\begingroup \@sanitize@url \@href}%
\providecommand \@href[1]{\@@startlink{#1}\@@href}%
\providecommand \@@href[1]{\endgroup#1\@@endlink}%
\providecommand \@sanitize@url [0]{\catcode `\\12\catcode `\$12\catcode
  `\&12\catcode `\#12\catcode `\^12\catcode `\_12\catcode `\%12\relax}%
\providecommand \@@startlink[1]{}%
\providecommand \@@endlink[0]{}%
\providecommand \url  [0]{\begingroup\@sanitize@url \@url }%
\providecommand \@url [1]{\endgroup\@href {#1}{\urlprefix }}%
\providecommand \urlprefix  [0]{URL }%
\providecommand \Eprint [0]{\href }%
\providecommand \doibase [0]{http://dx.doi.org/}%
\providecommand \selectlanguage [0]{\@gobble}%
\providecommand \bibinfo  [0]{\@secondoftwo}%
\providecommand \bibfield  [0]{\@secondoftwo}%
\providecommand \translation [1]{[#1]}%
\providecommand \BibitemOpen [0]{}%
\providecommand \bibitemStop [0]{}%
\providecommand \bibitemNoStop [0]{.\EOS\space}%
\providecommand \EOS [0]{\spacefactor3000\relax}%
\providecommand \BibitemShut  [1]{\csname bibitem#1\endcsname}%
\let\auto@bib@innerbib\@empty
\bibitem [{\citenamefont {Cyrot}\ \emph {et~al.}(1990)\citenamefont {Cyrot},
  \citenamefont {Lambert-Andron}, \citenamefont {Soubeyroux}, \citenamefont
  {Rey}, \citenamefont {Dehauht}, \citenamefont {Cyrot-Lackmann}, \citenamefont
  {Fourcaudot}, \citenamefont {Beille},\ and\ \citenamefont
  {Tholence}}]{Cyrot1990}%
  \BibitemOpen
  \bibfield  {author} {\bibinfo {author} {\bibfnamefont {M.}~\bibnamefont
  {Cyrot}}, \bibinfo {author} {\bibfnamefont {B.}~\bibnamefont
  {Lambert-Andron}}, \bibinfo {author} {\bibfnamefont {J.}~\bibnamefont
  {Soubeyroux}}, \bibinfo {author} {\bibfnamefont {M.}~\bibnamefont {Rey}},
  \bibinfo {author} {\bibfnamefont {P.}~\bibnamefont {Dehauht}}, \bibinfo
  {author} {\bibfnamefont {F.}~\bibnamefont {Cyrot-Lackmann}}, \bibinfo
  {author} {\bibfnamefont {G.}~\bibnamefont {Fourcaudot}}, \bibinfo {author}
  {\bibfnamefont {J.}~\bibnamefont {Beille}}, \ and\ \bibinfo {author}
  {\bibfnamefont {J.}~\bibnamefont {Tholence}},\ }\href {\doibase
  10.1016/s0022-4596(05)80090-8} {\bibfield  {journal} {\bibinfo  {journal}
  {Journal of Solid State Chemistry}\ }\textbf {\bibinfo {volume} {85}},\
  \bibinfo {pages} {321} (\bibinfo {year} {1990})}\BibitemShut {NoStop}%
\bibitem [{\citenamefont {Viennois}\ \emph {et~al.}(2010)\citenamefont
  {Viennois}, \citenamefont {Giannini}, \citenamefont {Teyssier}, \citenamefont
  {Elia}, \citenamefont {Deisenhofer},\ and\ \citenamefont {der
  Marel}}]{Viennois2010}%
  \BibitemOpen
  \bibfield  {author} {\bibinfo {author} {\bibfnamefont {R.}~\bibnamefont
  {Viennois}}, \bibinfo {author} {\bibfnamefont {E.}~\bibnamefont {Giannini}},
  \bibinfo {author} {\bibfnamefont {J.}~\bibnamefont {Teyssier}}, \bibinfo
  {author} {\bibfnamefont {J.}~\bibnamefont {Elia}}, \bibinfo {author}
  {\bibfnamefont {J.}~\bibnamefont {Deisenhofer}}, \ and\ \bibinfo {author}
  {\bibfnamefont {D.~V.}\ \bibnamefont {der Marel}},\ }\href {\doibase
  10.1088/1742-6596/200/1/012219} {\bibfield  {journal} {\bibinfo  {journal}
  {Journal of Physics: Conference Series}\ }\textbf {\bibinfo {volume} {200}},\
  \bibinfo {pages} {012219} (\bibinfo {year} {2010})}\BibitemShut {NoStop}%
\bibitem [{\citenamefont {Deslandes}\ \emph {et~al.}(1991)\citenamefont
  {Deslandes}, \citenamefont {Nazzal},\ and\ \citenamefont
  {Torrance}}]{Deslandes1991}%
  \BibitemOpen
  \bibfield  {author} {\bibinfo {author} {\bibfnamefont {F.}~\bibnamefont
  {Deslandes}}, \bibinfo {author} {\bibfnamefont {A.}~\bibnamefont {Nazzal}}, \
  and\ \bibinfo {author} {\bibfnamefont {J.}~\bibnamefont {Torrance}},\ }\href
  {\doibase 10.1016/0921-4534(91)90014-p} {\bibfield  {journal} {\bibinfo
  {journal} {Physica C: Superconductivity}\ }\textbf {\bibinfo {volume}
  {179}},\ \bibinfo {pages} {85} (\bibinfo {year} {1991})}\BibitemShut
  {NoStop}%
\bibitem [{\citenamefont {Wang}\ and\ \citenamefont
  {Senthil}(2011)}]{Wang2011}%
  \BibitemOpen
  \bibfield  {author} {\bibinfo {author} {\bibfnamefont {F.}~\bibnamefont
  {Wang}}\ and\ \bibinfo {author} {\bibfnamefont {T.}~\bibnamefont {Senthil}},\
  }\href {https://doi.org/10.1103/physrevlett.106.136402} {\bibfield  {journal}
  {\bibinfo  {journal} {Physical Review Letters}\ }\textbf {\bibinfo {volume}
  {106}} (\bibinfo {year} {2011})}\BibitemShut {NoStop}%
\bibitem [{\citenamefont {Wang}\ \emph {et~al.}(2018)\citenamefont {Wang},
  \citenamefont {Bachar}, \citenamefont {Teyssier}, \citenamefont {Luo},
  \citenamefont {Rischau}, \citenamefont {Scheerer}, \citenamefont {de~la
  Torre}, \citenamefont {Perry}, \citenamefont {Baumberger},\ and\
  \citenamefont {van~der Marel}}]{Wang2018}%
  \BibitemOpen
  \bibfield  {author} {\bibinfo {author} {\bibfnamefont {K.}~\bibnamefont
  {Wang}}, \bibinfo {author} {\bibfnamefont {N.}~\bibnamefont {Bachar}},
  \bibinfo {author} {\bibfnamefont {J.}~\bibnamefont {Teyssier}}, \bibinfo
  {author} {\bibfnamefont {W.}~\bibnamefont {Luo}}, \bibinfo {author}
  {\bibfnamefont {C.~W.}\ \bibnamefont {Rischau}}, \bibinfo {author}
  {\bibfnamefont {G.}~\bibnamefont {Scheerer}}, \bibinfo {author}
  {\bibfnamefont {A.}~\bibnamefont {de~la Torre}}, \bibinfo {author}
  {\bibfnamefont {R.~S.}\ \bibnamefont {Perry}}, \bibinfo {author}
  {\bibfnamefont {F.}~\bibnamefont {Baumberger}}, \ and\ \bibinfo {author}
  {\bibfnamefont {D.}~\bibnamefont {van~der Marel}},\ }\href
  {https://link.aps.org/doi/10.1103/PhysRevB.98.045107} {\bibfield  {journal}
  {\bibinfo  {journal} {Physical Review B}\ }\textbf {\bibinfo {volume} {98}}
  (\bibinfo {year} {2018})}\BibitemShut {NoStop}%
\bibitem [{\citenamefont {Chen}\ \emph {et~al.}(2015)\citenamefont {Chen},
  \citenamefont {Hogan}, \citenamefont {Walkup}, \citenamefont {Zhou},
  \citenamefont {Pokharel}, \citenamefont {Yao}, \citenamefont {Tian},
  \citenamefont {Ward}, \citenamefont {Zhao}, \citenamefont {Parshall},
  \citenamefont {Opeil}, \citenamefont {Lynn}, \citenamefont {Madhavan},\ and\
  \citenamefont {Wilson}}]{Chen2015}%
  \BibitemOpen
  \bibfield  {author} {\bibinfo {author} {\bibfnamefont {X.}~\bibnamefont
  {Chen}}, \bibinfo {author} {\bibfnamefont {T.}~\bibnamefont {Hogan}},
  \bibinfo {author} {\bibfnamefont {D.}~\bibnamefont {Walkup}}, \bibinfo
  {author} {\bibfnamefont {W.}~\bibnamefont {Zhou}}, \bibinfo {author}
  {\bibfnamefont {M.}~\bibnamefont {Pokharel}}, \bibinfo {author}
  {\bibfnamefont {M.}~\bibnamefont {Yao}}, \bibinfo {author} {\bibfnamefont
  {W.}~\bibnamefont {Tian}}, \bibinfo {author} {\bibfnamefont {T.~Z.}\
  \bibnamefont {Ward}}, \bibinfo {author} {\bibfnamefont {Y.}~\bibnamefont
  {Zhao}}, \bibinfo {author} {\bibfnamefont {D.}~\bibnamefont {Parshall}},
  \bibinfo {author} {\bibfnamefont {C.}~\bibnamefont {Opeil}}, \bibinfo
  {author} {\bibfnamefont {J.~W.}\ \bibnamefont {Lynn}}, \bibinfo {author}
  {\bibfnamefont {V.}~\bibnamefont {Madhavan}}, \ and\ \bibinfo {author}
  {\bibfnamefont {S.~D.}\ \bibnamefont {Wilson}},\ }\href
  {https://doi.org/10.1103/physrevb.92.075125} {\bibfield  {journal} {\bibinfo
  {journal} {Physical Review B}\ }\textbf {\bibinfo {volume} {92}} (\bibinfo
  {year} {2015})}\BibitemShut {NoStop}%
\bibitem [{\citenamefont {de~la Torre}\ \emph {et~al.}(2015)\citenamefont
  {de~la Torre}, \citenamefont {Walker}, \citenamefont {Bruno}, \citenamefont
  {Ricc{\'{o}}}, \citenamefont {Wang}, \citenamefont {Lezama}, \citenamefont
  {Scheerer}, \citenamefont {Giriat}, \citenamefont {Jaccard}, \citenamefont
  {Berthod}, \citenamefont {Kim}, \citenamefont {Hoesch}, \citenamefont
  {Hunter}, \citenamefont {Perry}, \citenamefont {Tamai},\ and\ \citenamefont
  {Baumberger}}]{delaTorre2015}%
  \BibitemOpen
  \bibfield  {author} {\bibinfo {author} {\bibfnamefont {A.}~\bibnamefont
  {de~la Torre}}, \bibinfo {author} {\bibfnamefont {S.~M.}\ \bibnamefont
  {Walker}}, \bibinfo {author} {\bibfnamefont {F.}~\bibnamefont {Bruno}},
  \bibinfo {author} {\bibfnamefont {S.}~\bibnamefont {Ricc{\'{o}}}}, \bibinfo
  {author} {\bibfnamefont {Z.}~\bibnamefont {Wang}}, \bibinfo {author}
  {\bibfnamefont {I.~G.}\ \bibnamefont {Lezama}}, \bibinfo {author}
  {\bibfnamefont {G.}~\bibnamefont {Scheerer}}, \bibinfo {author}
  {\bibfnamefont {G.}~\bibnamefont {Giriat}}, \bibinfo {author} {\bibfnamefont
  {D.}~\bibnamefont {Jaccard}}, \bibinfo {author} {\bibfnamefont
  {C.}~\bibnamefont {Berthod}}, \bibinfo {author} {\bibfnamefont
  {T.}~\bibnamefont {Kim}}, \bibinfo {author} {\bibfnamefont {M.}~\bibnamefont
  {Hoesch}}, \bibinfo {author} {\bibfnamefont {E.}~\bibnamefont {Hunter}},
  \bibinfo {author} {\bibfnamefont {R.}~\bibnamefont {Perry}}, \bibinfo
  {author} {\bibfnamefont {A.}~\bibnamefont {Tamai}}, \ and\ \bibinfo {author}
  {\bibfnamefont {F.}~\bibnamefont {Baumberger}},\ }\href
  {https://link.aps.org/doi/10.1103/PhysRevLett.115.176402} {\bibfield
  {journal} {\bibinfo  {journal} {Physical Review Letters}\ }\textbf {\bibinfo
  {volume} {115}} (\bibinfo {year} {2015})}\BibitemShut {NoStop}%
\bibitem [{\citenamefont {Li}\ \emph {et~al.}(2019)\citenamefont {Li},
  \citenamefont {Lee}, \citenamefont {Wang}, \citenamefont {Osada},
  \citenamefont {Crossley}, \citenamefont {Lee}, \citenamefont {Cui},
  \citenamefont {Hikita},\ and\ \citenamefont {Hwang}}]{Li2019}%
  \BibitemOpen
  \bibfield  {author} {\bibinfo {author} {\bibfnamefont {D.}~\bibnamefont
  {Li}}, \bibinfo {author} {\bibfnamefont {K.}~\bibnamefont {Lee}}, \bibinfo
  {author} {\bibfnamefont {B.~Y.}\ \bibnamefont {Wang}}, \bibinfo {author}
  {\bibfnamefont {M.}~\bibnamefont {Osada}}, \bibinfo {author} {\bibfnamefont
  {S.}~\bibnamefont {Crossley}}, \bibinfo {author} {\bibfnamefont {H.~R.}\
  \bibnamefont {Lee}}, \bibinfo {author} {\bibfnamefont {Y.}~\bibnamefont
  {Cui}}, \bibinfo {author} {\bibfnamefont {Y.}~\bibnamefont {Hikita}}, \ and\
  \bibinfo {author} {\bibfnamefont {H.~Y.}\ \bibnamefont {Hwang}},\ }\href
  {\doibase 10.1038/s41586-019-1496-5} {\bibfield  {journal} {\bibinfo
  {journal} {Nature}\ }\textbf {\bibinfo {volume} {572}},\ \bibinfo {pages}
  {624} (\bibinfo {year} {2019})}\BibitemShut {NoStop}%
\bibitem [{\citenamefont {Tjeng}\ \emph {et~al.}(1990)\citenamefont {Tjeng},
  \citenamefont {Meinders}, \citenamefont {van Elp}, \citenamefont {Ghijsen},
  \citenamefont {Sawatzky},\ and\ \citenamefont {Johnson}}]{Tjeng1990}%
  \BibitemOpen
  \bibfield  {author} {\bibinfo {author} {\bibfnamefont {L.~H.}\ \bibnamefont
  {Tjeng}}, \bibinfo {author} {\bibfnamefont {M.~B.~J.}\ \bibnamefont
  {Meinders}}, \bibinfo {author} {\bibfnamefont {J.}~\bibnamefont {van Elp}},
  \bibinfo {author} {\bibfnamefont {J.}~\bibnamefont {Ghijsen}}, \bibinfo
  {author} {\bibfnamefont {G.~A.}\ \bibnamefont {Sawatzky}}, \ and\ \bibinfo
  {author} {\bibfnamefont {R.~L.}\ \bibnamefont {Johnson}},\ }\href {\doibase
  10.1103/PhysRevB.41.3190} {\bibfield  {journal} {\bibinfo  {journal} {Phys.
  Rev. B}\ }\textbf {\bibinfo {volume} {41}},\ \bibinfo {pages} {3190}
  (\bibinfo {year} {1990})}\BibitemShut {NoStop}%
\bibitem [{\citenamefont {McLain}\ \emph {et~al.}(2006)\citenamefont {McLain},
  \citenamefont {Dolgos}, \citenamefont {Tennant}, \citenamefont {Turner},
  \citenamefont {Barnes}, \citenamefont {Proffen}, \citenamefont {Sales},\ and\
  \citenamefont {Bewley}}]{McLain2006}%
  \BibitemOpen
  \bibfield  {author} {\bibinfo {author} {\bibfnamefont {S.~E.}\ \bibnamefont
  {McLain}}, \bibinfo {author} {\bibfnamefont {M.~R.}\ \bibnamefont {Dolgos}},
  \bibinfo {author} {\bibfnamefont {D.~A.}\ \bibnamefont {Tennant}}, \bibinfo
  {author} {\bibfnamefont {J.~F.}\ \bibnamefont {Turner}}, \bibinfo {author}
  {\bibfnamefont {T.}~\bibnamefont {Barnes}}, \bibinfo {author} {\bibfnamefont
  {T.}~\bibnamefont {Proffen}}, \bibinfo {author} {\bibfnamefont {B.~C.}\
  \bibnamefont {Sales}}, \ and\ \bibinfo {author} {\bibfnamefont {R.~I.}\
  \bibnamefont {Bewley}},\ }\href {\doibase 10.1038/nmat1670} {\bibfield
  {journal} {\bibinfo  {journal} {Nat. Mater.}\ }\textbf {\bibinfo {volume}
  {5}},\ \bibinfo {pages} {561} (\bibinfo {year} {2006})}\BibitemShut {NoStop}%
\bibitem [{\citenamefont {Mazej}\ \emph {et~al.}(2009)\citenamefont {Mazej},
  \citenamefont {Goreshnik}, \citenamefont {Jagli{\v{c}}i{\'{c}}},
  \citenamefont {Gawe{\l}}, \citenamefont {{\L}asocha}, \citenamefont
  {Grzybowska}, \citenamefont {Jaro{\'{n}}}, \citenamefont {Kurzyd{\l}owski},
  \citenamefont {Malinowski}, \citenamefont {Ko{\'{z}}minski}, \citenamefont
  {Szyd{\l}owska}, \citenamefont {Leszczy{\'{n}}ski},\ and\ \citenamefont
  {Grochala}}]{Mazej2009}%
  \BibitemOpen
  \bibfield  {author} {\bibinfo {author} {\bibfnamefont {Z.}~\bibnamefont
  {Mazej}}, \bibinfo {author} {\bibfnamefont {E.}~\bibnamefont {Goreshnik}},
  \bibinfo {author} {\bibfnamefont {Z.}~\bibnamefont {Jagli{\v{c}}i{\'{c}}}},
  \bibinfo {author} {\bibfnamefont {B.}~\bibnamefont {Gawe{\l}}}, \bibinfo
  {author} {\bibfnamefont {W.}~\bibnamefont {{\L}asocha}}, \bibinfo {author}
  {\bibfnamefont {D.}~\bibnamefont {Grzybowska}}, \bibinfo {author}
  {\bibfnamefont {T.}~\bibnamefont {Jaro{\'{n}}}}, \bibinfo {author}
  {\bibfnamefont {D.}~\bibnamefont {Kurzyd{\l}owski}}, \bibinfo {author}
  {\bibfnamefont {P.}~\bibnamefont {Malinowski}}, \bibinfo {author}
  {\bibfnamefont {W.}~\bibnamefont {Ko{\'{z}}minski}}, \bibinfo {author}
  {\bibfnamefont {J.}~\bibnamefont {Szyd{\l}owska}}, \bibinfo {author}
  {\bibfnamefont {P.}~\bibnamefont {Leszczy{\'{n}}ski}}, \ and\ \bibinfo
  {author} {\bibfnamefont {W.}~\bibnamefont {Grochala}},\ }\href {\doibase
  10.1039/b902161b} {\bibfield  {journal} {\bibinfo  {journal} {CrystEngComm}\
  }\textbf {\bibinfo {volume} {11}},\ \bibinfo {pages} {1702} (\bibinfo {year}
  {2009})}\BibitemShut {NoStop}%
\bibitem [{\citenamefont {Grochala}(2006)}]{Grochala2006}%
  \BibitemOpen
  \bibfield  {author} {\bibinfo {author} {\bibfnamefont {W.}~\bibnamefont
  {Grochala}},\ }\href {\doibase 10.1038/nmat1678} {\bibfield  {journal}
  {\bibinfo  {journal} {Nat. Mater.}\ }\textbf {\bibinfo {volume} {5}},\
  \bibinfo {pages} {513} (\bibinfo {year} {2006})}\BibitemShut {NoStop}%
\bibitem [{\citenamefont {Kurzyd{\l}owski}\ \emph {et~al.}(2013)\citenamefont
  {Kurzyd{\l}owski}, \citenamefont {Mazej}, \citenamefont
  {Jagli{\v{c}}i{\'{c}}}, \citenamefont {Filinchuk},\ and\ \citenamefont
  {Grochala}}]{Kurzydowski2013}%
  \BibitemOpen
  \bibfield  {author} {\bibinfo {author} {\bibfnamefont {D.}~\bibnamefont
  {Kurzyd{\l}owski}}, \bibinfo {author} {\bibfnamefont {Z.}~\bibnamefont
  {Mazej}}, \bibinfo {author} {\bibfnamefont {Z.}~\bibnamefont
  {Jagli{\v{c}}i{\'{c}}}}, \bibinfo {author} {\bibfnamefont {Y.}~\bibnamefont
  {Filinchuk}}, \ and\ \bibinfo {author} {\bibfnamefont {W.}~\bibnamefont
  {Grochala}},\ }\href {\doibase 10.1039/c3cc41521j} {\bibfield  {journal}
  {\bibinfo  {journal} {Chem. Commun.}\ }\textbf {\bibinfo {volume} {49}},\
  \bibinfo {pages} {6262} (\bibinfo {year} {2013})}\BibitemShut {NoStop}%
\bibitem [{\citenamefont {Gawraczy{\'n}ski}\ \emph {et~al.}(2019)\citenamefont
  {Gawraczy{\'n}ski}, \citenamefont {Kurzyd{\l}owski}, \citenamefont {Ewings},
  \citenamefont {Bandaru}, \citenamefont {Gadomski}, \citenamefont {Mazej},
  \citenamefont {Ruani}, \citenamefont {Bergenti}, \citenamefont {Jaro{\'n}},
  \citenamefont {Ozarowski}, \citenamefont {Hill}, \citenamefont
  {Leszczy{\'n}ski}, \citenamefont {Tok{\'a}r}, \citenamefont {Derzsi},
  \citenamefont {Barone}, \citenamefont {Wohlfeld}, \citenamefont {Lorenzana},\
  and\ \citenamefont {Grochala}}]{Gawraczynski2019}%
  \BibitemOpen
  \bibfield  {author} {\bibinfo {author} {\bibfnamefont {J.}~\bibnamefont
  {Gawraczy{\'n}ski}}, \bibinfo {author} {\bibfnamefont {D.}~\bibnamefont
  {Kurzyd{\l}owski}}, \bibinfo {author} {\bibfnamefont {R.~A.}\ \bibnamefont
  {Ewings}}, \bibinfo {author} {\bibfnamefont {S.}~\bibnamefont {Bandaru}},
  \bibinfo {author} {\bibfnamefont {W.}~\bibnamefont {Gadomski}}, \bibinfo
  {author} {\bibfnamefont {Z.}~\bibnamefont {Mazej}}, \bibinfo {author}
  {\bibfnamefont {G.}~\bibnamefont {Ruani}}, \bibinfo {author} {\bibfnamefont
  {I.}~\bibnamefont {Bergenti}}, \bibinfo {author} {\bibfnamefont
  {T.}~\bibnamefont {Jaro{\'n}}}, \bibinfo {author} {\bibfnamefont
  {A.}~\bibnamefont {Ozarowski}}, \bibinfo {author} {\bibfnamefont
  {S.}~\bibnamefont {Hill}}, \bibinfo {author} {\bibfnamefont {P.~J.}\
  \bibnamefont {Leszczy{\'n}ski}}, \bibinfo {author} {\bibfnamefont
  {K.}~\bibnamefont {Tok{\'a}r}}, \bibinfo {author} {\bibfnamefont
  {M.}~\bibnamefont {Derzsi}}, \bibinfo {author} {\bibfnamefont
  {P.}~\bibnamefont {Barone}}, \bibinfo {author} {\bibfnamefont
  {K.}~\bibnamefont {Wohlfeld}}, \bibinfo {author} {\bibfnamefont
  {J.}~\bibnamefont {Lorenzana}}, \ and\ \bibinfo {author} {\bibfnamefont
  {W.}~\bibnamefont {Grochala}},\ }\href {\doibase 10.1073/pnas.1812857116}
  {\bibfield  {journal} {\bibinfo  {journal} {Proceedings of the National
  Academy of Sciences}\ }\textbf {\bibinfo {volume} {116}},\ \bibinfo {pages}
  {1495} (\bibinfo {year} {2019})}\BibitemShut {NoStop}%
\bibitem [{\citenamefont {Jaro{\'{n}}}\ and\ \citenamefont
  {Grochala}(2008)}]{Jaron2008}%
  \BibitemOpen
  \bibfield  {author} {\bibinfo {author} {\bibfnamefont {T.}~\bibnamefont
  {Jaro{\'{n}}}}\ and\ \bibinfo {author} {\bibfnamefont {W.}~\bibnamefont
  {Grochala}},\ }\href {\doibase 10.1002/pssr.200701286} {\bibfield  {journal}
  {\bibinfo  {journal} {Phys. status solidi – Rapid Res. Lett.}\ }\textbf
  {\bibinfo {volume} {2}},\ \bibinfo {pages} {71} (\bibinfo {year}
  {2008})}\BibitemShut {NoStop}%
\bibitem [{\citenamefont {Grzelak}\ \emph {et~al.}(2017)\citenamefont
  {Grzelak}, \citenamefont {Gawraczy{\'{n}}ski}, \citenamefont {Jaro{\'{n}}},
  \citenamefont {Kurzyd{\l}owski}, \citenamefont {Budzianowski}, \citenamefont
  {Mazej}, \citenamefont {Leszczy{\'{n}}ski}, \citenamefont {Prakapenka},
  \citenamefont {Derzsi}, \citenamefont {Struzhkin},\ and\ \citenamefont
  {Grochala}}]{Grzelak2017}%
  \BibitemOpen
  \bibfield  {author} {\bibinfo {author} {\bibfnamefont {A.}~\bibnamefont
  {Grzelak}}, \bibinfo {author} {\bibfnamefont {J.}~\bibnamefont
  {Gawraczy{\'{n}}ski}}, \bibinfo {author} {\bibfnamefont {T.}~\bibnamefont
  {Jaro{\'{n}}}}, \bibinfo {author} {\bibfnamefont {D.}~\bibnamefont
  {Kurzyd{\l}owski}}, \bibinfo {author} {\bibfnamefont {A.}~\bibnamefont
  {Budzianowski}}, \bibinfo {author} {\bibfnamefont {Z.}~\bibnamefont {Mazej}},
  \bibinfo {author} {\bibfnamefont {P.~J.}\ \bibnamefont {Leszczy{\'{n}}ski}},
  \bibinfo {author} {\bibfnamefont {V.~B.}\ \bibnamefont {Prakapenka}},
  \bibinfo {author} {\bibfnamefont {M.}~\bibnamefont {Derzsi}}, \bibinfo
  {author} {\bibfnamefont {V.~V.}\ \bibnamefont {Struzhkin}}, \ and\ \bibinfo
  {author} {\bibfnamefont {W.}~\bibnamefont {Grochala}},\ }\href {\doibase
  10.1021/acs.inorgchem.7b02528} {\bibfield  {journal} {\bibinfo  {journal}
  {Inorg. Chem.}\ }\textbf {\bibinfo {volume} {56}},\ \bibinfo {pages} {14651}
  (\bibinfo {year} {2017})}\BibitemShut {NoStop}%
\bibitem [{\citenamefont {Kurzyd{\l}owski}\ and\ \citenamefont
  {Grochala}(2017)}]{Kurzydowski2017}%
  \BibitemOpen
  \bibfield  {author} {\bibinfo {author} {\bibfnamefont {D.}~\bibnamefont
  {Kurzyd{\l}owski}}\ and\ \bibinfo {author} {\bibfnamefont {W.}~\bibnamefont
  {Grochala}},\ }\href {\doibase 10.1002/anie.201700932} {\bibfield  {journal}
  {\bibinfo  {journal} {Angew. Chemie - Int. Ed.}\ }\textbf {\bibinfo {volume}
  {56}},\ \bibinfo {pages} {10114} (\bibinfo {year} {2017})}\BibitemShut
  {NoStop}%
\bibitem [{RIX()}]{RIXS_Diamond}%
  \BibitemOpen
  \href {https://www.diamond.ac.uk/Instruments/Magnetic-Materials/I21.html}
  {\enquote {\bibinfo {title} {Diamond {RIXS I21}},}\ }\BibitemShut {NoStop}%
\bibitem [{\citenamefont {Pelliciari}\ \emph {et~al.}(2016)\citenamefont
  {Pelliciari}, \citenamefont {Huang}, \citenamefont {Das}, \citenamefont
  {Dantz}, \citenamefont {Bisogni}, \citenamefont {Velasco}, \citenamefont
  {Strocov}, \citenamefont {Xing}, \citenamefont {Wang}, \citenamefont {Jin},\
  and\ \citenamefont {Schmitt}}]{pelliciari2016}%
  \BibitemOpen
  \bibfield  {author} {\bibinfo {author} {\bibfnamefont {J.}~\bibnamefont
  {Pelliciari}}, \bibinfo {author} {\bibfnamefont {Y.}~\bibnamefont {Huang}},
  \bibinfo {author} {\bibfnamefont {T.}~\bibnamefont {Das}}, \bibinfo {author}
  {\bibfnamefont {M.}~\bibnamefont {Dantz}}, \bibinfo {author} {\bibfnamefont
  {V.}~\bibnamefont {Bisogni}}, \bibinfo {author} {\bibfnamefont {P.~O.}\
  \bibnamefont {Velasco}}, \bibinfo {author} {\bibfnamefont {V.~N.}\
  \bibnamefont {Strocov}}, \bibinfo {author} {\bibfnamefont {L.}~\bibnamefont
  {Xing}}, \bibinfo {author} {\bibfnamefont {X.}~\bibnamefont {Wang}}, \bibinfo
  {author} {\bibfnamefont {C.}~\bibnamefont {Jin}}, \ and\ \bibinfo {author}
  {\bibfnamefont {T.}~\bibnamefont {Schmitt}},\ }\href {\doibase
  10.1103/PhysRevB.93.134515} {\bibfield  {journal} {\bibinfo  {journal} {Phys.
  Rev. B}\ }\textbf {\bibinfo {volume} {93}},\ \bibinfo {pages} {134515}
  (\bibinfo {year} {2016})}\BibitemShut {NoStop}%
\bibitem [{\citenamefont {Kresse}\ and\ \citenamefont
  {Furthm{\"{u}}ller}(1996)}]{Kresse1996}%
  \BibitemOpen
  \bibfield  {author} {\bibinfo {author} {\bibfnamefont {G.}~\bibnamefont
  {Kresse}}\ and\ \bibinfo {author} {\bibfnamefont {J.}~\bibnamefont
  {Furthm{\"{u}}ller}},\ }\href {\doibase 10.1103/PhysRevB.54.11169} {\bibfield
   {journal} {\bibinfo  {journal} {Phys. Rev. B}\ }\textbf {\bibinfo {volume}
  {54}},\ \bibinfo {pages} {11169} (\bibinfo {year} {1996})}\BibitemShut
  {NoStop}%
\bibitem [{\citenamefont {Perdew}\ \emph {et~al.}(1996)\citenamefont {Perdew},
  \citenamefont {Burke},\ and\ \citenamefont {Ernzerhof}}]{Perdew1996}%
  \BibitemOpen
  \bibfield  {author} {\bibinfo {author} {\bibfnamefont {J.~P.}\ \bibnamefont
  {Perdew}}, \bibinfo {author} {\bibfnamefont {K.}~\bibnamefont {Burke}}, \
  and\ \bibinfo {author} {\bibfnamefont {M.}~\bibnamefont {Ernzerhof}},\ }\href
  {\doibase 10.1103/PhysRevLett.77.3865} {\bibfield  {journal} {\bibinfo
  {journal} {Phys. Rev. Lett.}\ }\textbf {\bibinfo {volume} {77}},\ \bibinfo
  {pages} {3865} (\bibinfo {year} {1996})}\BibitemShut {NoStop}%
\bibitem [{\citenamefont {Fischer}\ \emph {et~al.}(1971)\citenamefont
  {Fischer}, \citenamefont {Roult},\ and\ \citenamefont
  {Schwarzenbach}}]{Fischer1971}%
  \BibitemOpen
  \bibfield  {author} {\bibinfo {author} {\bibfnamefont {P.}~\bibnamefont
  {Fischer}}, \bibinfo {author} {\bibfnamefont {G.}~\bibnamefont {Roult}}, \
  and\ \bibinfo {author} {\bibfnamefont {D.}~\bibnamefont {Schwarzenbach}},\
  }\href {\doibase 10.1016/S0022-3697(71)80057-4} {\bibfield  {journal}
  {\bibinfo  {journal} {J. Phys. Chem. Solids}\ }\textbf {\bibinfo {volume}
  {32}},\ \bibinfo {pages} {1641} (\bibinfo {year} {1971})}\BibitemShut
  {NoStop}%
\bibitem [{\citenamefont {Marzari}\ \emph {et~al.}(2012)\citenamefont
  {Marzari}, \citenamefont {Mostofi}, \citenamefont {Yates}, \citenamefont
  {Souza},\ and\ \citenamefont {Vanderbilt}}]{Marzari2012}%
  \BibitemOpen
  \bibfield  {author} {\bibinfo {author} {\bibfnamefont {N.}~\bibnamefont
  {Marzari}}, \bibinfo {author} {\bibfnamefont {A.~a.}\ \bibnamefont
  {Mostofi}}, \bibinfo {author} {\bibfnamefont {J.~R.}\ \bibnamefont {Yates}},
  \bibinfo {author} {\bibfnamefont {I.}~\bibnamefont {Souza}}, \ and\ \bibinfo
  {author} {\bibfnamefont {D.}~\bibnamefont {Vanderbilt}},\ }\href {\doibase
  10.1103/RevModPhys.84.1419} {\bibfield  {journal} {\bibinfo  {journal} {Rev.
  Mod. Phys.}\ }\textbf {\bibinfo {volume} {84}},\ \bibinfo {pages} {1419}
  (\bibinfo {year} {2012})}\BibitemShut {NoStop}%
\bibitem [{\citenamefont {Eskes}\ \emph {et~al.}(1990)\citenamefont {Eskes},
  \citenamefont {Tjeng},\ and\ \citenamefont {Sawatzky}}]{Eskes1990}%
  \BibitemOpen
  \bibfield  {author} {\bibinfo {author} {\bibfnamefont {H.}~\bibnamefont
  {Eskes}}, \bibinfo {author} {\bibfnamefont {L.~H.}\ \bibnamefont {Tjeng}}, \
  and\ \bibinfo {author} {\bibfnamefont {G.~A.}\ \bibnamefont {Sawatzky}},\
  }\href {\doibase 10.1103/PhysRevB.41.288} {\bibfield  {journal} {\bibinfo
  {journal} {Phys. Rev. B}\ }\textbf {\bibinfo {volume} {41}},\ \bibinfo
  {pages} {288} (\bibinfo {year} {1990})}\BibitemShut {NoStop}%
\bibitem [{\citenamefont {Falck}\ \emph {et~al.}(1992)\citenamefont {Falck},
  \citenamefont {Levy}, \citenamefont {Kastner},\ and\ \citenamefont
  {Birgeneau}}]{Falck1992}%
  \BibitemOpen
  \bibfield  {author} {\bibinfo {author} {\bibfnamefont {J.~P.}\ \bibnamefont
  {Falck}}, \bibinfo {author} {\bibfnamefont {A.}~\bibnamefont {Levy}},
  \bibinfo {author} {\bibfnamefont {M.~A.}\ \bibnamefont {Kastner}}, \ and\
  \bibinfo {author} {\bibfnamefont {R.~J.}\ \bibnamefont {Birgeneau}},\ }\href
  {\doibase 10.1103/PhysRevLett.69.1109} {\bibfield  {journal} {\bibinfo
  {journal} {Phys. Rev. Lett.}\ }\textbf {\bibinfo {volume} {69}},\ \bibinfo
  {pages} {1109} (\bibinfo {year} {1992})}\BibitemShut {NoStop}%
\bibitem [{\citenamefont {Falck}\ \emph {et~al.}(1994)\citenamefont {Falck},
  \citenamefont {Perkins}, \citenamefont {Levy}, \citenamefont {Kastner},
  \citenamefont {Graybeal},\ and\ \citenamefont {Birgeneau}}]{Falck1994}%
  \BibitemOpen
  \bibfield  {author} {\bibinfo {author} {\bibfnamefont {J.~P.}\ \bibnamefont
  {Falck}}, \bibinfo {author} {\bibfnamefont {J.~D.}\ \bibnamefont {Perkins}},
  \bibinfo {author} {\bibfnamefont {A.}~\bibnamefont {Levy}}, \bibinfo {author}
  {\bibfnamefont {M.~A.}\ \bibnamefont {Kastner}}, \bibinfo {author}
  {\bibfnamefont {J.~M.}\ \bibnamefont {Graybeal}}, \ and\ \bibinfo {author}
  {\bibfnamefont {R.~J.}\ \bibnamefont {Birgeneau}},\ }\href {\doibase
  10.1103/PhysRevB.49.6246} {\bibfield  {journal} {\bibinfo  {journal} {Phys.
  Rev. B}\ }\textbf {\bibinfo {volume} {49}},\ \bibinfo {pages} {6246}
  (\bibinfo {year} {1994})}\BibitemShut {NoStop}%
\bibitem [{\citenamefont {Chen}\ \emph {et~al.}(1991)\citenamefont {Chen},
  \citenamefont {Sette}, \citenamefont {Ma}, \citenamefont {Hybertsen},
  \citenamefont {Stechel}, \citenamefont {Foulkes}, \citenamefont {Schulter},
  \citenamefont {Cheong}, \citenamefont {Cooper}, \citenamefont {Rupp},
  \citenamefont {Batlogg}, \citenamefont {Soo}, \citenamefont {Ming},
  \citenamefont {Krol},\ and\ \citenamefont {Kao}}]{chen1994}%
  \BibitemOpen
  \bibfield  {author} {\bibinfo {author} {\bibfnamefont {C.~T.}\ \bibnamefont
  {Chen}}, \bibinfo {author} {\bibfnamefont {F.}~\bibnamefont {Sette}},
  \bibinfo {author} {\bibfnamefont {Y.}~\bibnamefont {Ma}}, \bibinfo {author}
  {\bibfnamefont {M.~S.}\ \bibnamefont {Hybertsen}}, \bibinfo {author}
  {\bibfnamefont {E.~B.}\ \bibnamefont {Stechel}}, \bibinfo {author}
  {\bibfnamefont {W.~M.~C.}\ \bibnamefont {Foulkes}}, \bibinfo {author}
  {\bibfnamefont {M.}~\bibnamefont {Schulter}}, \bibinfo {author}
  {\bibfnamefont {S.-W.}\ \bibnamefont {Cheong}}, \bibinfo {author}
  {\bibfnamefont {A.~S.}\ \bibnamefont {Cooper}}, \bibinfo {author}
  {\bibfnamefont {L.~W.}\ \bibnamefont {Rupp}}, \bibinfo {author}
  {\bibfnamefont {B.}~\bibnamefont {Batlogg}}, \bibinfo {author} {\bibfnamefont
  {Y.~L.}\ \bibnamefont {Soo}}, \bibinfo {author} {\bibfnamefont {Z.~H.}\
  \bibnamefont {Ming}}, \bibinfo {author} {\bibfnamefont {A.}~\bibnamefont
  {Krol}}, \ and\ \bibinfo {author} {\bibfnamefont {Y.~H.}\ \bibnamefont
  {Kao}},\ }\href {\doibase 10.1103/PhysRevLett.66.104} {\bibfield  {journal}
  {\bibinfo  {journal} {Phys. Rev. Lett.}\ }\textbf {\bibinfo {volume} {66}},\
  \bibinfo {pages} {104} (\bibinfo {year} {1991})}\BibitemShut {NoStop}%
\bibitem [{\citenamefont {Nakai}\ \emph {et~al.}(1988)\citenamefont {Nakai},
  \citenamefont {Kawata}, \citenamefont {Ohashi}, \citenamefont {Kitamura},
  \citenamefont {Sugiura}, \citenamefont {Mitsuishi},\ and\ \citenamefont
  {Maezawa}}]{nakai1988}%
  \BibitemOpen
  \bibfield  {author} {\bibinfo {author} {\bibfnamefont {S.}~\bibnamefont
  {Nakai}}, \bibinfo {author} {\bibfnamefont {A.}~\bibnamefont {Kawata}},
  \bibinfo {author} {\bibfnamefont {M.}~\bibnamefont {Ohashi}}, \bibinfo
  {author} {\bibfnamefont {M.}~\bibnamefont {Kitamura}}, \bibinfo {author}
  {\bibfnamefont {C.}~\bibnamefont {Sugiura}}, \bibinfo {author} {\bibfnamefont
  {T.}~\bibnamefont {Mitsuishi}}, \ and\ \bibinfo {author} {\bibfnamefont
  {H.}~\bibnamefont {Maezawa}},\ }\href {\doibase 10.1103/PhysRevB.37.10895}
  {\bibfield  {journal} {\bibinfo  {journal} {Phys. Rev. B}\ }\textbf {\bibinfo
  {volume} {37}},\ \bibinfo {pages} {10895} (\bibinfo {year}
  {1988})}\BibitemShut {NoStop}%
\bibitem [{\citenamefont {Olalde-Velasco}\ \emph {et~al.}(2013)\citenamefont
  {Olalde-Velasco}, \citenamefont {Jim\'enez-Mier}, \citenamefont {Denlinger},\
  and\ \citenamefont {Yang}}]{velasco2013}%
  \BibitemOpen
  \bibfield  {author} {\bibinfo {author} {\bibfnamefont {P.}~\bibnamefont
  {Olalde-Velasco}}, \bibinfo {author} {\bibfnamefont {J.}~\bibnamefont
  {Jim\'enez-Mier}}, \bibinfo {author} {\bibfnamefont {J.}~\bibnamefont
  {Denlinger}}, \ and\ \bibinfo {author} {\bibfnamefont {W.-L.}\ \bibnamefont
  {Yang}},\ }\href {\doibase 10.1103/PhysRevB.87.245136} {\bibfield  {journal}
  {\bibinfo  {journal} {Phys. Rev. B}\ }\textbf {\bibinfo {volume} {87}},\
  \bibinfo {pages} {245136} (\bibinfo {year} {2013})}\BibitemShut {NoStop}%
\bibitem [{\citenamefont {Bondino}\ \emph {et~al.}(2009)\citenamefont
  {Bondino}, \citenamefont {Malvestuto}, \citenamefont {Magnano}, \citenamefont
  {Zangrando}, \citenamefont {Zacchigna}, \citenamefont {Ghigna},\ and\
  \citenamefont {Parmigiani}}]{bondino2009}%
  \BibitemOpen
  \bibfield  {author} {\bibinfo {author} {\bibfnamefont {F.}~\bibnamefont
  {Bondino}}, \bibinfo {author} {\bibfnamefont {M.}~\bibnamefont {Malvestuto}},
  \bibinfo {author} {\bibfnamefont {E.}~\bibnamefont {Magnano}}, \bibinfo
  {author} {\bibfnamefont {M.}~\bibnamefont {Zangrando}}, \bibinfo {author}
  {\bibfnamefont {M.}~\bibnamefont {Zacchigna}}, \bibinfo {author}
  {\bibfnamefont {P.}~\bibnamefont {Ghigna}}, \ and\ \bibinfo {author}
  {\bibfnamefont {F.}~\bibnamefont {Parmigiani}},\ }\href {\doibase
  10.1103/PhysRevB.79.115120} {\bibfield  {journal} {\bibinfo  {journal} {Phys.
  Rev. B}\ }\textbf {\bibinfo {volume} {79}},\ \bibinfo {pages} {115120}
  (\bibinfo {year} {2009})}\BibitemShut {NoStop}%
\bibitem [{\citenamefont {Bisogni}\ \emph {et~al.}(2012)\citenamefont
  {Bisogni}, \citenamefont {Simonelli}, \citenamefont {Ament}, \citenamefont
  {Forte}, \citenamefont {Moretti~Sala}, \citenamefont {Minola}, \citenamefont
  {Huotari}, \citenamefont {van~den Brink}, \citenamefont {Ghiringhelli},
  \citenamefont {Brookes},\ and\ \citenamefont {Braicovich}}]{valentina2012}%
  \BibitemOpen
  \bibfield  {author} {\bibinfo {author} {\bibfnamefont {V.}~\bibnamefont
  {Bisogni}}, \bibinfo {author} {\bibfnamefont {L.}~\bibnamefont {Simonelli}},
  \bibinfo {author} {\bibfnamefont {L.~J.~P.}\ \bibnamefont {Ament}}, \bibinfo
  {author} {\bibfnamefont {F.}~\bibnamefont {Forte}}, \bibinfo {author}
  {\bibfnamefont {M.}~\bibnamefont {Moretti~Sala}}, \bibinfo {author}
  {\bibfnamefont {M.}~\bibnamefont {Minola}}, \bibinfo {author} {\bibfnamefont
  {S.}~\bibnamefont {Huotari}}, \bibinfo {author} {\bibfnamefont
  {J.}~\bibnamefont {van~den Brink}}, \bibinfo {author} {\bibfnamefont
  {G.}~\bibnamefont {Ghiringhelli}}, \bibinfo {author} {\bibfnamefont {N.~B.}\
  \bibnamefont {Brookes}}, \ and\ \bibinfo {author} {\bibfnamefont
  {L.}~\bibnamefont {Braicovich}},\ }\href {\doibase
  10.1103/PhysRevB.85.214527} {\bibfield  {journal} {\bibinfo  {journal} {Phys.
  Rev. B}\ }\textbf {\bibinfo {volume} {85}},\ \bibinfo {pages} {214527}
  (\bibinfo {year} {2012})}\BibitemShut {NoStop}%
\bibitem [{\citenamefont {Uchida}\ \emph {et~al.}(1991)\citenamefont {Uchida},
  \citenamefont {Ido}, \citenamefont {Takagi}, \citenamefont {Arima},
  \citenamefont {Tokura},\ and\ \citenamefont {Tajima}}]{Uchida1991}%
  \BibitemOpen
  \bibfield  {author} {\bibinfo {author} {\bibfnamefont {S.}~\bibnamefont
  {Uchida}}, \bibinfo {author} {\bibfnamefont {T.}~\bibnamefont {Ido}},
  \bibinfo {author} {\bibfnamefont {H.}~\bibnamefont {Takagi}}, \bibinfo
  {author} {\bibfnamefont {T.}~\bibnamefont {Arima}}, \bibinfo {author}
  {\bibfnamefont {Y.}~\bibnamefont {Tokura}}, \ and\ \bibinfo {author}
  {\bibfnamefont {S.}~\bibnamefont {Tajima}},\ }\href {\doibase
  10.1103/PhysRevB.43.7942} {\bibfield  {journal} {\bibinfo  {journal} {Phys.
  Rev. B}\ }\textbf {\bibinfo {volume} {43}},\ \bibinfo {pages} {7942}
  (\bibinfo {year} {1991})}\BibitemShut {NoStop}%
\bibitem [{\citenamefont {Thomas}\ \emph {et~al.}(1992)\citenamefont {Thomas},
  \citenamefont {Rapkine}, \citenamefont {Cooper}, \citenamefont {Cheong},
  \citenamefont {Cooper}, \citenamefont {Schneemeyer},\ and\ \citenamefont
  {Waszczak}}]{Thomas1992}%
  \BibitemOpen
  \bibfield  {author} {\bibinfo {author} {\bibfnamefont {G.~A.}\ \bibnamefont
  {Thomas}}, \bibinfo {author} {\bibfnamefont {D.~H.}\ \bibnamefont {Rapkine}},
  \bibinfo {author} {\bibfnamefont {S.~L.}\ \bibnamefont {Cooper}}, \bibinfo
  {author} {\bibfnamefont {S.-W.}\ \bibnamefont {Cheong}}, \bibinfo {author}
  {\bibfnamefont {A.~S.}\ \bibnamefont {Cooper}}, \bibinfo {author}
  {\bibfnamefont {L.~F.}\ \bibnamefont {Schneemeyer}}, \ and\ \bibinfo {author}
  {\bibfnamefont {J.~V.}\ \bibnamefont {Waszczak}},\ }\href {\doibase
  10.1103/PhysRevB.45.2474} {\bibfield  {journal} {\bibinfo  {journal} {Phys.
  Rev. B}\ }\textbf {\bibinfo {volume} {45}},\ \bibinfo {pages} {2474}
  (\bibinfo {year} {1992})}\BibitemShut {NoStop}%
\bibitem [{\citenamefont {Cooper}\ \emph {et~al.}(1990)\citenamefont {Cooper},
  \citenamefont {Thomas}, \citenamefont {Millis}, \citenamefont {Sulewski},
  \citenamefont {Orenstein}, \citenamefont {Rapkine}, \citenamefont {Cheong},\
  and\ \citenamefont {Trevor}}]{Cooper1990}%
  \BibitemOpen
  \bibfield  {author} {\bibinfo {author} {\bibfnamefont {S.~L.}\ \bibnamefont
  {Cooper}}, \bibinfo {author} {\bibfnamefont {G.~A.}\ \bibnamefont {Thomas}},
  \bibinfo {author} {\bibfnamefont {A.~J.}\ \bibnamefont {Millis}}, \bibinfo
  {author} {\bibfnamefont {P.~E.}\ \bibnamefont {Sulewski}}, \bibinfo {author}
  {\bibfnamefont {J.}~\bibnamefont {Orenstein}}, \bibinfo {author}
  {\bibfnamefont {D.~H.}\ \bibnamefont {Rapkine}}, \bibinfo {author}
  {\bibfnamefont {S.-W.}\ \bibnamefont {Cheong}}, \ and\ \bibinfo {author}
  {\bibfnamefont {P.~L.}\ \bibnamefont {Trevor}},\ }\href {\doibase
  10.1103/PhysRevB.42.10785} {\bibfield  {journal} {\bibinfo  {journal} {Phys.
  Rev. B}\ }\textbf {\bibinfo {volume} {42}},\ \bibinfo {pages} {10785}
  (\bibinfo {year} {1990})}\BibitemShut {NoStop}%
\bibitem [{\citenamefont {Perkins}\ \emph {et~al.}(1993)\citenamefont
  {Perkins}, \citenamefont {Graybeal}, \citenamefont {Kastner}, \citenamefont
  {Birgeneau}, \citenamefont {Falck},\ and\ \citenamefont
  {Greven}}]{Perkins1993}%
  \BibitemOpen
  \bibfield  {author} {\bibinfo {author} {\bibfnamefont {J.~D.}\ \bibnamefont
  {Perkins}}, \bibinfo {author} {\bibfnamefont {J.~M.}\ \bibnamefont
  {Graybeal}}, \bibinfo {author} {\bibfnamefont {M.~A.}\ \bibnamefont
  {Kastner}}, \bibinfo {author} {\bibfnamefont {R.~J.}\ \bibnamefont
  {Birgeneau}}, \bibinfo {author} {\bibfnamefont {J.~P.}\ \bibnamefont
  {Falck}}, \ and\ \bibinfo {author} {\bibfnamefont {M.}~\bibnamefont
  {Greven}},\ }\href {\doibase 10.1103/PhysRevLett.71.1621} {\bibfield
  {journal} {\bibinfo  {journal} {Phys. Rev. Lett.}\ }\textbf {\bibinfo
  {volume} {71}},\ \bibinfo {pages} {1621} (\bibinfo {year}
  {1993})}\BibitemShut {NoStop}%
\bibitem [{\citenamefont {Basov}\ and\ \citenamefont
  {Timusk}(2005)}]{Basov2005}%
  \BibitemOpen
  \bibfield  {author} {\bibinfo {author} {\bibfnamefont {D.~N.}\ \bibnamefont
  {Basov}}\ and\ \bibinfo {author} {\bibfnamefont {T.}~\bibnamefont {Timusk}},\
  }\href {\doibase 10.1103/RevModPhys.77.721} {\bibfield  {journal} {\bibinfo
  {journal} {Rev. Mod. Phys.}\ }\textbf {\bibinfo {volume} {77}},\ \bibinfo
  {pages} {721} (\bibinfo {year} {2005})}\BibitemShut {NoStop}%
\bibitem [{\citenamefont {Tajima}(2016)}]{Tajima2016}%
  \BibitemOpen
  \bibfield  {author} {\bibinfo {author} {\bibfnamefont {S.}~\bibnamefont
  {Tajima}},\ }\href {\doibase 10.1088/0034-4885/79/9/094001} {\bibfield
  {journal} {\bibinfo  {journal} {Reports on Progress in Physics}\ }\textbf
  {\bibinfo {volume} {79}},\ \bibinfo {pages} {094001} (\bibinfo {year}
  {2016})}\BibitemShut {NoStop}%
\bibitem [{\citenamefont {Friebel}\ and\ \citenamefont
  {Reinen}(1975)}]{Friebel1975}%
  \BibitemOpen
  \bibfield  {author} {\bibinfo {author} {\bibfnamefont {C.}~\bibnamefont
  {Friebel}}\ and\ \bibinfo {author} {\bibfnamefont {D.}~\bibnamefont
  {Reinen}},\ }\href {\doibase 10.1002/zaac.19754130107} {\bibfield  {journal}
  {\bibinfo  {journal} {Zeitschrift f\"{u}r Anorg. und Allg. Chemie}\ }\textbf
  {\bibinfo {volume} {413}},\ \bibinfo {pages} {51} (\bibinfo {year}
  {1975})}\BibitemShut {NoStop}%
\bibitem [{\citenamefont {Monnier}\ \emph {et~al.}(1991)\citenamefont
  {Monnier}, \citenamefont {Gerber},\ and\ \citenamefont {Bill}}]{Monnier1991}%
  \BibitemOpen
  \bibfield  {author} {\bibinfo {author} {\bibfnamefont {A.}~\bibnamefont
  {Monnier}}, \bibinfo {author} {\bibfnamefont {A.}~\bibnamefont {Gerber}}, \
  and\ \bibinfo {author} {\bibfnamefont {H.}~\bibnamefont {Bill}},\ }\href
  {\doibase 10.1063/1.460473} {\bibfield  {journal} {\bibinfo  {journal} {J.
  Chem. Phys.}\ }\textbf {\bibinfo {volume} {94}},\ \bibinfo {pages} {5891}
  (\bibinfo {year} {1991})}\BibitemShut {NoStop}%
\bibitem [{\citenamefont {Aramburu}\ \emph {et~al.}(1992)\citenamefont
  {Aramburu}, \citenamefont {Moreno},\ and\ \citenamefont
  {Barriuso}}]{Aramburu1992}%
  \BibitemOpen
  \bibfield  {author} {\bibinfo {author} {\bibfnamefont {J.~A.}\ \bibnamefont
  {Aramburu}}, \bibinfo {author} {\bibfnamefont {M.}~\bibnamefont {Moreno}}, \
  and\ \bibinfo {author} {\bibfnamefont {M.~T.}\ \bibnamefont {Barriuso}},\
  }\href {\doibase 10.1088/0953-8984/4/46/016} {\bibfield  {journal} {\bibinfo
  {journal} {J. Phys. Condens. Matter}\ }\textbf {\bibinfo {volume} {4}},\
  \bibinfo {pages} {9089} (\bibinfo {year} {1992})}\BibitemShut {NoStop}%
\bibitem [{\citenamefont {Valiente}\ \emph {et~al.}(1994)\citenamefont
  {Valiente}, \citenamefont {Aramburu}, \citenamefont {Barriuso},\ and\
  \citenamefont {Moreno}}]{Valiente1994}%
  \BibitemOpen
  \bibfield  {author} {\bibinfo {author} {\bibfnamefont {R.}~\bibnamefont
  {Valiente}}, \bibinfo {author} {\bibfnamefont {J.~A.}\ \bibnamefont
  {Aramburu}}, \bibinfo {author} {\bibfnamefont {M.~T.}\ \bibnamefont
  {Barriuso}}, \ and\ \bibinfo {author} {\bibfnamefont {M.}~\bibnamefont
  {Moreno}},\ }\href {\doibase 10.1088/0953-8984/6/24/013} {\bibfield
  {journal} {\bibinfo  {journal} {J. Phys. Condens. Matter}\ }\textbf {\bibinfo
  {volume} {6}},\ \bibinfo {pages} {4515} (\bibinfo {year} {1994})}\BibitemShut
  {NoStop}%
\bibitem [{\citenamefont {Mazej}\ \emph {et~al.}(2015)\citenamefont {Mazej},
  \citenamefont {Micha{\l}owski}, \citenamefont {Goreshnik}, \citenamefont
  {Jagli{\v{c}}i{\'{c}}}, \citenamefont {Ar{\v{c}}on}, \citenamefont
  {Szyd{\l}owska},\ and\ \citenamefont {Grochala}}]{Mazej2015}%
  \BibitemOpen
  \bibfield  {author} {\bibinfo {author} {\bibfnamefont {Z.}~\bibnamefont
  {Mazej}}, \bibinfo {author} {\bibfnamefont {T.}~\bibnamefont
  {Micha{\l}owski}}, \bibinfo {author} {\bibfnamefont {E.~A.}\ \bibnamefont
  {Goreshnik}}, \bibinfo {author} {\bibfnamefont {Z.}~\bibnamefont
  {Jagli{\v{c}}i{\'{c}}}}, \bibinfo {author} {\bibfnamefont {I.}~\bibnamefont
  {Ar{\v{c}}on}}, \bibinfo {author} {\bibfnamefont {J.}~\bibnamefont
  {Szyd{\l}owska}}, \ and\ \bibinfo {author} {\bibfnamefont {W.}~\bibnamefont
  {Grochala}},\ }\href {\doibase 10.1039/c5dt00740b} {\bibfield  {journal}
  {\bibinfo  {journal} {Dalt. Trans.}\ }\textbf {\bibinfo {volume} {44}},\
  \bibinfo {pages} {10957} (\bibinfo {year} {2015})}\BibitemShut {NoStop}%
\bibitem [{\citenamefont {Anisimov}\ \emph {et~al.}(1992)\citenamefont
  {Anisimov}, \citenamefont {Korotin}, \citenamefont {Zaanen},\ and\
  \citenamefont {Andersen}}]{Anisimov1992}%
  \BibitemOpen
  \bibfield  {author} {\bibinfo {author} {\bibfnamefont {V.~I.}\ \bibnamefont
  {Anisimov}}, \bibinfo {author} {\bibfnamefont {M.~A.}\ \bibnamefont
  {Korotin}}, \bibinfo {author} {\bibfnamefont {J.}~\bibnamefont {Zaanen}}, \
  and\ \bibinfo {author} {\bibfnamefont {O.~K.}\ \bibnamefont {Andersen}},\
  }\href {\doibase 10.1103/PhysRevLett.68.345} {\bibfield  {journal} {\bibinfo
  {journal} {Phys. Rev. Lett.}\ }\textbf {\bibinfo {volume} {68}},\ \bibinfo
  {pages} {345} (\bibinfo {year} {1992})}\BibitemShut {NoStop}%
\bibitem [{\citenamefont {Hozoi}\ \emph {et~al.}(2011)\citenamefont {Hozoi},
  \citenamefont {Siurakshina}, \citenamefont {Fulde},\ and\ \citenamefont {{Van
  Den Brink}}}]{Hozoi2011}%
  \BibitemOpen
  \bibfield  {author} {\bibinfo {author} {\bibfnamefont {L.}~\bibnamefont
  {Hozoi}}, \bibinfo {author} {\bibfnamefont {L.}~\bibnamefont {Siurakshina}},
  \bibinfo {author} {\bibfnamefont {P.}~\bibnamefont {Fulde}}, \ and\ \bibinfo
  {author} {\bibfnamefont {J.}~\bibnamefont {{Van Den Brink}}},\ }\href
  {\doibase 10.1038/srep00065} {\bibfield  {journal} {\bibinfo  {journal} {Sci.
  Rep.}\ }\textbf {\bibinfo {volume} {1}},\ \bibinfo {pages} {1} (\bibinfo
  {year} {2011})}\BibitemShut {NoStop}%
\bibitem [{\citenamefont {Sala}\ \emph {et~al.}(2011)\citenamefont {Sala},
  \citenamefont {Bisogni}, \citenamefont {Aruta}, \citenamefont {Balestrino},
  \citenamefont {Berger}, \citenamefont {Brookes}, \citenamefont {de~Luca},
  \citenamefont {Castro}, \citenamefont {Grioni}, \citenamefont {Guarise},
  \citenamefont {Medaglia}, \citenamefont {Granozio}, \citenamefont {Minola},
  \citenamefont {Perna}, \citenamefont {Radovic}, \citenamefont {Salluzzo},
  \citenamefont {Schmitt}, \citenamefont {Zhou}, \citenamefont {Braicovich},\
  and\ \citenamefont {Ghiringhelli}}]{MorettiSala2011}%
  \BibitemOpen
  \bibfield  {author} {\bibinfo {author} {\bibfnamefont {M.~M.}\ \bibnamefont
  {Sala}}, \bibinfo {author} {\bibfnamefont {V.}~\bibnamefont {Bisogni}},
  \bibinfo {author} {\bibfnamefont {C.}~\bibnamefont {Aruta}}, \bibinfo
  {author} {\bibfnamefont {G.}~\bibnamefont {Balestrino}}, \bibinfo {author}
  {\bibfnamefont {H.}~\bibnamefont {Berger}}, \bibinfo {author} {\bibfnamefont
  {N.~B.}\ \bibnamefont {Brookes}}, \bibinfo {author} {\bibfnamefont {G.~M.}\
  \bibnamefont {de~Luca}}, \bibinfo {author} {\bibfnamefont {D.~D.}\
  \bibnamefont {Castro}}, \bibinfo {author} {\bibfnamefont {M.}~\bibnamefont
  {Grioni}}, \bibinfo {author} {\bibfnamefont {M.}~\bibnamefont {Guarise}},
  \bibinfo {author} {\bibfnamefont {P.~G.}\ \bibnamefont {Medaglia}}, \bibinfo
  {author} {\bibfnamefont {F.~M.}\ \bibnamefont {Granozio}}, \bibinfo {author}
  {\bibfnamefont {M.}~\bibnamefont {Minola}}, \bibinfo {author} {\bibfnamefont
  {P.}~\bibnamefont {Perna}}, \bibinfo {author} {\bibfnamefont
  {M.}~\bibnamefont {Radovic}}, \bibinfo {author} {\bibfnamefont
  {M.}~\bibnamefont {Salluzzo}}, \bibinfo {author} {\bibfnamefont
  {T.}~\bibnamefont {Schmitt}}, \bibinfo {author} {\bibfnamefont {K.~J.}\
  \bibnamefont {Zhou}}, \bibinfo {author} {\bibfnamefont {L.}~\bibnamefont
  {Braicovich}}, \ and\ \bibinfo {author} {\bibfnamefont {G.}~\bibnamefont
  {Ghiringhelli}},\ }\href {\doibase 10.1088/1367-2630/13/4/043026} {\bibfield
  {journal} {\bibinfo  {journal} {New Journal of Physics}\ }\textbf {\bibinfo
  {volume} {13}},\ \bibinfo {pages} {043026} (\bibinfo {year}
  {2011})}\BibitemShut {NoStop}%
\bibitem [{\citenamefont {Dagotto}(1994)}]{Dagotto1994}%
  \BibitemOpen
  \bibfield  {author} {\bibinfo {author} {\bibfnamefont {E.}~\bibnamefont
  {Dagotto}},\ }\href {\doibase 10.1103/RevModPhys.66.763} {\bibfield
  {journal} {\bibinfo  {journal} {Rev. Mod. Phys.}\ }\textbf {\bibinfo {volume}
  {66}},\ \bibinfo {pages} {763} (\bibinfo {year} {1994})},\ \Eprint
  {http://arxiv.org/abs/9311013} {9311013} \BibitemShut {NoStop}%
\bibitem [{\citenamefont {Haverkort}\ \emph {et~al.}(2012)\citenamefont
  {Haverkort}, \citenamefont {Zwierzycki},\ and\ \citenamefont
  {Andersen}}]{Haverkort2012}%
  \BibitemOpen
  \bibfield  {author} {\bibinfo {author} {\bibfnamefont {M.~W.}\ \bibnamefont
  {Haverkort}}, \bibinfo {author} {\bibfnamefont {M.}~\bibnamefont
  {Zwierzycki}}, \ and\ \bibinfo {author} {\bibfnamefont {O.~K.}\ \bibnamefont
  {Andersen}},\ }\href {\doibase 10.1103/PhysRevB.85.165113} {\bibfield
  {journal} {\bibinfo  {journal} {Phys. Rev. B}\ }\textbf {\bibinfo {volume}
  {85}},\ \bibinfo {pages} {165113} (\bibinfo {year} {2012})}\BibitemShut
  {NoStop}%
\bibitem [{\citenamefont {Kohn}\ and\ \citenamefont {Sham}(1965)}]{Kohn1965}%
  \BibitemOpen
  \bibfield  {author} {\bibinfo {author} {\bibfnamefont {W.}~\bibnamefont
  {Kohn}}\ and\ \bibinfo {author} {\bibfnamefont {L.~J.}\ \bibnamefont
  {Sham}},\ }\href {\doibase 10.1103/PhysRev.140.A1133} {\bibfield  {journal}
  {\bibinfo  {journal} {Phys. Rev.}\ }\textbf {\bibinfo {volume} {140}},\
  \bibinfo {pages} {A1133} (\bibinfo {year} {1965})}\BibitemShut {NoStop}%
\bibitem [{\citenamefont {Brosco}\ \emph {et~al.}(2013)\citenamefont {Brosco},
  \citenamefont {Ying},\ and\ \citenamefont {Lorenzana}}]{Brosco2013}%
  \BibitemOpen
  \bibfield  {author} {\bibinfo {author} {\bibfnamefont {V.}~\bibnamefont
  {Brosco}}, \bibinfo {author} {\bibfnamefont {Z.-J.}\ \bibnamefont {Ying}}, \
  and\ \bibinfo {author} {\bibfnamefont {J.}~\bibnamefont {Lorenzana}},\
  }\href@noop {} {\bibfield  {journal} {\bibinfo  {journal} {Sci. Rep.}\
  }\textbf {\bibinfo {volume} {3}} (\bibinfo {year} {2013})}\BibitemShut
  {NoStop}%
\bibitem [{\citenamefont {Grochala}\ \emph {et~al.}(2003)\citenamefont
  {Grochala}, \citenamefont {Egdell}, \citenamefont {Edwards}, \citenamefont
  {Mazej},\ and\ \citenamefont {{\v{Z}}emva}}]{Grochala2003}%
  \BibitemOpen
  \bibfield  {author} {\bibinfo {author} {\bibfnamefont {W.}~\bibnamefont
  {Grochala}}, \bibinfo {author} {\bibfnamefont {R.~G.}\ \bibnamefont
  {Egdell}}, \bibinfo {author} {\bibfnamefont {P.~P.}\ \bibnamefont {Edwards}},
  \bibinfo {author} {\bibfnamefont {Z.}~\bibnamefont {Mazej}}, \ and\ \bibinfo
  {author} {\bibfnamefont {B.}~\bibnamefont {{\v{Z}}emva}},\ }\href {\doibase
  10.1002/cphc.200300777} {\bibfield  {journal} {\bibinfo  {journal}
  {{ChemPhysChem}}\ }\textbf {\bibinfo {volume} {4}},\ \bibinfo {pages} {997}
  (\bibinfo {year} {2003})}\BibitemShut {NoStop}%
\bibitem [{\citenamefont {Shen}\ \emph {et~al.}(1999)\citenamefont {Shen},
  \citenamefont {{\v{Z}}emva}, \citenamefont {Lucier}, \citenamefont
  {Graudejus}, \citenamefont {Allman},\ and\ \citenamefont
  {Bartlett}}]{Shen1999}%
  \BibitemOpen
  \bibfield  {author} {\bibinfo {author} {\bibfnamefont {C.}~\bibnamefont
  {Shen}}, \bibinfo {author} {\bibfnamefont {B.}~\bibnamefont {{\v{Z}}emva}},
  \bibinfo {author} {\bibfnamefont {G.~M.}\ \bibnamefont {Lucier}}, \bibinfo
  {author} {\bibfnamefont {O.}~\bibnamefont {Graudejus}}, \bibinfo {author}
  {\bibfnamefont {J.~A.}\ \bibnamefont {Allman}}, \ and\ \bibinfo {author}
  {\bibfnamefont {N.}~\bibnamefont {Bartlett}},\ }\href {\doibase
  10.1021/ic9905603} {\bibfield  {journal} {\bibinfo  {journal} {Inorg. Chem.}\
  }\textbf {\bibinfo {volume} {38}},\ \bibinfo {pages} {4570} (\bibinfo {year}
  {1999})}\BibitemShut {NoStop}%
\bibitem [{\citenamefont {Romiszewski}\ \emph {et~al.}(2007)\citenamefont
  {Romiszewski}, \citenamefont {Grochala},\ and\ \citenamefont
  {Stolarczyk}}]{Romiszewski2007}%
  \BibitemOpen
  \bibfield  {author} {\bibinfo {author} {\bibfnamefont {J.}~\bibnamefont
  {Romiszewski}}, \bibinfo {author} {\bibfnamefont {W.}~\bibnamefont
  {Grochala}}, \ and\ \bibinfo {author} {\bibfnamefont {L.~Z.}\ \bibnamefont
  {Stolarczyk}},\ }\href {\doibase 10.1088/0953-8984/19/11/116206} {\bibfield
  {journal} {\bibinfo  {journal} {Journal of Physics: Condensed Matter}\
  }\textbf {\bibinfo {volume} {19}},\ \bibinfo {pages} {116206} (\bibinfo
  {year} {2007})}\BibitemShut {NoStop}%
\bibitem [{\citenamefont {Tokár}\ \emph {et~al.}(2021)\citenamefont {Tokár},
  \citenamefont {Derzsi},\ and\ \citenamefont {Grochala}}]{Tokar2021}%
  \BibitemOpen
  \bibfield  {author} {\bibinfo {author} {\bibfnamefont {K.}~\bibnamefont
  {Tokár}}, \bibinfo {author} {\bibfnamefont {M.}~\bibnamefont {Derzsi}}, \
  and\ \bibinfo {author} {\bibfnamefont {W.}~\bibnamefont {Grochala}},\ }\href
  {\doibase https://doi.org/10.1016/j.commatsci.2020.110250} {\bibfield
  {journal} {\bibinfo  {journal} {Computational Materials Science}\ }\textbf
  {\bibinfo {volume} {188}},\ \bibinfo {pages} {110250} (\bibinfo {year}
  {2021})}\BibitemShut {NoStop}%
\bibitem [{\citenamefont {Varma}(2012)}]{Varma2012}%
  \BibitemOpen
  \bibfield  {author} {\bibinfo {author} {\bibfnamefont {C.~M.}\ \bibnamefont
  {Varma}},\ }\href {\doibase 10.1088/0034-4885/75/5/052501} {\bibfield
  {journal} {\bibinfo  {journal} {Reports Prog. Phys.}\ }\textbf {\bibinfo
  {volume} {75}},\ \bibinfo {pages} {052501} (\bibinfo {year}
  {2012})}\BibitemShut {NoStop}%
\bibitem [{\citenamefont {Bandaru}\ \emph {et~al.}(2021)\citenamefont
  {Bandaru}, \citenamefont {Derzsi}, \citenamefont {Grzelak}, \citenamefont
  {Lorenzana},\ and\ \citenamefont {Grochala}}]{Bandaru2021}%
  \BibitemOpen
  \bibfield  {author} {\bibinfo {author} {\bibfnamefont {S.}~\bibnamefont
  {Bandaru}}, \bibinfo {author} {\bibfnamefont {M.}~\bibnamefont {Derzsi}},
  \bibinfo {author} {\bibfnamefont {A.}~\bibnamefont {Grzelak}}, \bibinfo
  {author} {\bibfnamefont {J.}~\bibnamefont {Lorenzana}}, \ and\ \bibinfo
  {author} {\bibfnamefont {W.}~\bibnamefont {Grochala}},\ }\href {\doibase
  10.1103/PhysRevMaterials.5.064801} {\bibfield  {journal} {\bibinfo  {journal}
  {Phys. Rev. Mater.}\ }\textbf {\bibinfo {volume} {5}},\ \bibinfo {pages}
  {064801} (\bibinfo {year} {2021})}\BibitemShut {NoStop}%
\bibitem [{\citenamefont {Grzelak}\ \emph {et~al.}(2020)\citenamefont
  {Grzelak}, \citenamefont {Su}, \citenamefont {Yang}, \citenamefont
  {Kurzyd{\l}owski}, \citenamefont {Lorenzana},\ and\ \citenamefont
  {Grochala}}]{Grzelak2020}%
  \BibitemOpen
  \bibfield  {author} {\bibinfo {author} {\bibfnamefont {A.}~\bibnamefont
  {Grzelak}}, \bibinfo {author} {\bibfnamefont {H.}~\bibnamefont {Su}},
  \bibinfo {author} {\bibfnamefont {X.}~\bibnamefont {Yang}}, \bibinfo {author}
  {\bibfnamefont {D.}~\bibnamefont {Kurzyd{\l}owski}}, \bibinfo {author}
  {\bibfnamefont {J.}~\bibnamefont {Lorenzana}}, \ and\ \bibinfo {author}
  {\bibfnamefont {W.}~\bibnamefont {Grochala}},\ }\href {\doibase
  10.1103/physrevmaterials.4.084405} {\bibfield  {journal} {\bibinfo  {journal}
  {Phys. Rev. Mater.}\ }\textbf {\bibinfo {volume} {4}},\ \bibinfo {pages}
  {084405} (\bibinfo {year} {2020})},\ \Eprint
  {http://arxiv.org/abs/2005.00461} {arXiv:2005.00461} \BibitemShut {NoStop}%
\bibitem [{\citenamefont {Trębi\'{n}ski}\ \emph {et~al.}(1986)\citenamefont
  {Trębi\'{n}ski}, \citenamefont {Trzci\'{n}ski},\ and\ \citenamefont
  {W{\l}odarczyk}}]{Trebinski1986}%
  \BibitemOpen
  \bibfield  {author} {\bibinfo {author} {\bibfnamefont {R.}~\bibnamefont
  {Trębi\'{n}ski}}, \bibinfo {author} {\bibfnamefont {W.~A.}\ \bibnamefont
  {Trzci\'{n}ski}}, \ and\ \bibinfo {author} {\bibfnamefont {E.}~\bibnamefont
  {W{\l}odarczyk}},\ }\href@noop {} {\bibfield  {journal} {\bibinfo  {journal}
  {J. Techn. Phys.}\ }\textbf {\bibinfo {volume} {27}},\ \bibinfo {pages} {3}
  (\bibinfo {year} {1986})}\BibitemShut {NoStop}%
\end{thebibliography}%

\end{document}